\documentclass[usenatbib]{mn2e}
\pdfoutput=1
\usepackage{amssymb}
\usepackage{amsfonts}
\usepackage{amsmath}
\usepackage[pdftex]{graphicx}
\usepackage{epsfig}
\usepackage{color}
\usepackage{dblfloatfix}
\usepackage{float}
\usepackage{fixltx2e}
\usepackage{hyperref}
\usepackage{url}
\def\twid{\mathrel{\lower.1ex\hbox{$\sim$}}}
\def\gtwid{\mathrel{\raise.3ex\hbox{$>$\kern-.75em\lower1ex\hbox{$\sim$}}}}
\def\ltwid{\mathrel{\raise.3ex\hbox{$<$\kern-.75em\lower1ex\hbox{$\sim$}}}}
\def\\{\hfil\break}
\def\ie{{\it i.e.\ }}
\def\eg{{\it e.g.\ }}

\def\hmpc{$h^{-1}$Mpc}
\def\kmSecMpc{km s$^{-1}$ Mpc$^{-1}$}

\newcommand{\be}{\begin{equation}}
\newcommand{\ee}{\end{equation}}
\newcommand{\bea}{\begin{eqnarray}}
\newcommand{\eea}{\end{eqnarray}}

\begin{document}

\title[PkANN--Matter power spectrum interpolator]{PkANN -- II. A non-linear matter power spectrum interpolator developed using artificial neural networks}

\author[Agarwal, Abdalla, Feldman, Lahav \& Thomas]{\Large
Shankar Agarwal$^{1, \dagger}$, Filipe B. Abdalla$^{3, \S}$, Hume A. Feldman$^{2, \ddagger}$, Ofer Lahav$^{3, \flat}$ \& Shaun A. Thomas$^{3, \star}$\\
$^1$CNRS, Laboratoire Univers et Th\'eories (LUTh), UMR 8102 CNRS, Observatoire de Paris, Universit\'e Paris Diderot,\\ \,\;5 Place Jules Janssen, 92190 Meudon, France.\\
$^2$Department of Physics \& Astronomy, University of Kansas, Lawrence, KS 66045, USA.\\
$^3$Department of Physics \& Astronomy, University College London, Gower Street, London, WC1E 6BT, UK.\\
emails: $^{\dagger}$shankar.agarwal@obspm.fr, $^{\S}$fba@star.ucl.ac.uk, $^{\ddagger}$feldman@ku.edu, $^{\flat}$lahav@star.ucl.ac.uk, $^{\star}$sat@star.ucl.ac.uk
}

\date{}

\maketitle

\begin{abstract}
In this paper we introduce {\sc PkANN}, a freely available software package for interpolating the non-linear matter power spectrum, constructed using Artificial Neural Networks (ANNs). Previously, using {\sc halofit} to calculate matter power spectrum, we demonstrated that ANNs can make extremely quick and accurate predictions of the power spectrum. Now, using a suite of 6380 N-body simulations spanning 580 cosmologies, we train ANNs to predict the power spectrum over the cosmological parameter space spanning $3\sigma$ confidence level (CL) around the concordance cosmology. When presented with a set of cosmological parameters ($\Omega_{\rm m} h^2, \Omega_{\rm b} h^2, n_s, w, \sigma_8, \sum m_\nu$ and redshift $z$), the trained ANN interpolates the power spectrum for $z\leq2$ at sub-per cent accuracy for modes up to $k\leq0.9\,h\textrm{Mpc}^{-1}$. {\sc PkANN} is faster than computationally expensive N-body simulations, yet provides a worst-case error $<1$ per cent fit to the non-linear matter power spectrum deduced through N-body simulations. The overall precision of {\sc PkANN} is set by the accuracy of our N-body simulations, at 5 per cent level for cosmological models with $\sum m_\nu<0.5$ eV for all redshifts $z\leq2$. For models with $\sum m_\nu>0.5$ eV, predictions are expected to be at 5 (10) per cent level for redshifts $z>1$ ($z\leq1$). The {\sc PkANN} interpolator may be freely downloaded from \url{http://zuserver2.star.ucl.ac.uk/~fba/PkANN}.

\end{abstract}


\section{Introduction}

With the upcoming surveys promising to breach the per cent level of precision, any efforts to further improve the constraints on cosmological parameters will be predominantly theory limited. The Baryon Oscillation Spectroscopic Survey (BOSS) \citep{BOSS11} aims to determine the angular diameter distance with a precision of 1 per cent at redshifts $z = 0.3$ and $z = 0.55$, and the cosmic expansion rate $H(z)$ with 1-2 per cent precision at the same redshifts. The DES \citep{DES05} will probe the nature of dark energy through both the growth of structure in the universe as a function of time and the dependence of distances on the expansion rate. The Dark Energy Spectroscopic Instrument (DESI) \citep{DESI2013}, through redshift measurements of millions of galaxies and quasars, will enable baryon acoustic oscillation (BAO) and redshift space distortion measurements. The Large Synoptic Survey Telescope (LSST) \citep{LSST08}, LSST will measure the comoving distance in the redshift range $z=0.3-3.0$ with an accuracy of 1-2 per cent. These studies will shed more light and possibly solve some of the unanswered questions in cosmology including the nature of dark energy, and the absolute mass scale, the hierarchy and the effective number of neutrino species. Using BAO and the cosmic microwave background (CMB) data, Planck \citep{Planck13} constrains the dark energy constant equation of state parameter at $w=-1.13\pm0.13$ with no evidence for dynamical dark energy. This is consistent with a cosmological constant ($w=-1$) dominated flat universe. In order to distinguish between various models of dark energy, such as $w\ne-1$ and/or a time-varying equation of state parameter, one needs more precise and accurate measurements of the matter power spectrum.

Neutrino oscillation experiments \citep{MINOS,SNO,KamLAND} indicate that at least two neutrino eigentstates have non-zero masses. Massive neutrinos thus qualify as a hot dark matter component and contribute to the total energy density of the Universe. Free-streaming of massive neutrinos damps small-scale density perturbations, thereby suppressing the growth of cosmological structure. Accurate measurements of the matter power spectrum offer a powerful tool to constrain the absolute mass-scale of neutrinos, and complement the oscillation experiments which, being sensitive to the mass squared differences between the neutrino eigentstates, only provide a lower bound on the total neutrino mass. Specifically, mass splittings of $|\Delta m^{2}_{32}|=(2.43\pm0.13)\times10^{-3}\,{\rm eV}^2$ and $\Delta m^{2}_{21}=(7.59\pm0.21)\times10^{-5}\,{\rm eV}^2$ \citep{MINOS,KamLAND} imply a lower limit for the sum of the neutrino masses to be $0.06$ and $0.1\,{\rm eV}$ for the normal and inverted mass hierarchies \citep{OttenWein08}, respectively. Assuming a minimal-mass ($\sum m_\nu=0.06$ eV) normal hierarchy for the neutrino masses, the Planck survey find $\sum m_\nu<0.23$ eV (95\% CL). WMAP 9-year \citep{WMAP9} analysis find $\sum m_\nu<0.44$ eV (95\% CL). \cite{LahKiaAbdBla2010} obtained an upper limit of 0.11 eV (95\% CL). Numerical studies of the scale-dependent suppression of matter power spectrum has been performed by various groups: \cite{BraHan10}; \cite*{VieHaeSpr10,AgaFel10,BirdVielHae12,WagVerJim12}. \cite*{AgaFel10} (hereafter, Paper I) and \cite*{WagVerJim12} show that resolving the neutrino mass hierarchy may require the power spectrum to be measured at better than $0.5$ per cent accuracy, which may be possible with the next generation of experiments.

Currently, there are four popular approaches to estimate the non-linear matter power spectrum: (i) {\sc halofit} \citep{Smith03}, (ii) higher order perturbation theory (PT, \eg \citealt{Saito08}; \citealt{NisShiTarYah09}; \citealt*{Saito09}; \citealt{UpaBisPopHeiHabFInFro13}), (iii) N-body simulations (\eg {\sc enzo} \citealt{OShea10} and {\sc gadget} \citealt{Springel05}); (iv) spectrum interpolators (\eg \citealt{Heit06}; \citealt{Habib07}; \citealt{Lawrence10}; \citealt{Heitmann13}). While {\sc halofit} performs well on large scales ($k\ltwid0.1\,h \textrm{Mpc}^{-1}$), its performance degrades rapidly on smaller scales. \cite{TakSatNisTarOgu12} re-calibrated the original {\sc halofit} \citep{Smith03} extending it to include dark energy models with constant equation of state $w\ne-1$. The accuracy of {\sc halofit} predictions is model dependent and may be as low as $5-10$ per cent at $k\sim1\,h \textrm{Mpc}^{-1}$ (\citealt{TakSatNisTarOgu12,Heitmann13}). Likewise, PT improves upon linear theory predictions on large scales but fails on smaller ($k\!\gtwid\!0.09\,h \textrm{Mpc}^{-1}$) scales. At higher redshifts when the perturbations are less evolved, the accuracy for both {\sc halofit} and PT improves. However, since dark energy is a late-time phenomenon ($z\ltwid2$), one can not rely on fitting functions like {\sc halofit} and PT at low redshifts if one aims to develop a theoretical framework capable of predicting the non-linear matter power spectrum at per cent level. This leaves N-body simulations as the only method capable of controlling the accuracy levels as desired. \cite{Heitmann10} show that gravity-only simulations can be used to calculate the matter power spectrum at sub-per cent accuracy up to $k\ltwid1\,h \textrm{Mpc}^{-1}$. On smaller scales, baryonic physics affects the power spectrum and needs to be included in numerical simulations to maintain per cent accuracy.

A typical high-resolution dark-matter only simulation intended to probe $k\ltwid1\,h \textrm{Mpc}^{-1}$ scales can cost $\sim10,000$ CPU-hours. Including hydrodynamics in simulations to probe smaller scales can take prohibitively long, especially when running multiple simulations spread across the cosmological parameter space. As discussed earlier in \cite{Heit06} and \cite{Habib07}, parameter estimation and model building typically involves sampling the parameter space and evaluating the power spectrum for each cosmology. As we mentioned in \cite{AgaAbdFelLahTho12} (hereafter, Paper II), given the multi-dimensionality of the cosmological parameter space, a brute force application of N-body simulations is beyond our current state of the art computing capabilities.

A novel alternative to running numerical simulations to determine the non-linear response from varying parameter settings, is to use Machine-learning techniques. Machine-learning has found use in a variety of applications such as brain-machine interfaces \citep{JenGraMicOboEgeThi11,PedGraVarThi12}, analyses of stock market \citep{Ghosh11, HurMar12}, fitting of cosmological functions (\citealt{Auld07}; \citealt{Fendt07}; \citealt*{Auld08}), and estimating photometric redshifts \citep{Collister04}.

Using Machine-learning in the form of Gaussian processes \cite{Heitmann09, Lawrence10, Heitmann13} have developed a matter power spectrum calculator -- {\sc cosmic emulator}, that is an order of magnitude improvement over the popular {\sc halofit} prescription. The {\sc cosmic emulator}, based on gravity-only N-body simulations, comes in two versions: $h$-fixed \citep{Lawrence10} and $h$-free \citep{Heitmann13}. The $h$-fixed version computes the Hubble parameter $h$ using the CMB constraint on the acoustic scale and predicts the non-linear matter power spectrum up to $z\leq1$ for modes $k\ltwid1\,h \textrm{Mpc}^{-1}$. The $h$-free version has $h$ as a free parameter that can be set by the user. The range of validity of the $h$-free version is up to $z\leq4$ for modes $k\ltwid15\,h \textrm{Mpc}^{-1}$. Both versions are restricted to cosmological models with massless neutrinos. Since the current understanding is that at least two neutrino eigentstates have non-zero masses, it is reasonable to develop a power spectrum interpolator that is suitable for cosmological models with/without massive neutrinos.

In paper II, we developed the formalism for estimating the non-linear matter power spectrum using ANNs. Using {\sc halofit} spectra as mock N-body spectra, we showed that the ANN formalism enables a remarkable fit with a manageable number of simulations. In this paper, we use a suite of 6380 N-body simulations spanning 580 cosmologies around the WMAP 7-year central values, and train ANNs to predict the power spectrum accurate at 5-10 per cent level for $k\leq0.9\,h \textrm{Mpc}^{-1}$ up to redshifts $z\leq2$. The {\sc PkANN} package, along with instructions to use, is available at \url{http://zuserver2.star.ucl.ac.uk/~fba/PkANN}.

We trained {\sc PkANN} for a range of cosmologies including $w\ne-1$ and $m_\nu \ne0$. However, the training can be easily extended to include other parameters such as time-varying dark energy, modified gravity as well as probing small scale baryonic effects. This will require (i) running a few N-body simulations around the cosmological parameter(s) being probed; (ii) calculating the matter power spectra from numerical simulations; (iii) randomly dividing these power spectra into two sets, namely, the training and validation sets (explained in Paper II, and here in Appendix~\ref{sec:PkANN}); (iv) training {\sc PkANN} using the training and validation sets. Once training is over, the trained network can be used to predict the matter power spectrum at new parameter settings.  

The outline of this paper is as follows. We discuss our numerical simulations in Section~\ref{sec:SIM}. We develop the {\sc PkANN} interpolator in Section~\ref{sec:ANN}. We present our results in Section~\ref{sec:RESULTS} starting with the performance of the {\sc PkANN} interpolator against spectra computed using N-body simulations (Section~\ref{sec:PkANN_vs_SIM}). The estimate of errors in {\sc PkANN}'s predictions are summarized in Section~\ref{sec:ERRORS}. In Section~\ref{sec:EXPLORE}, we use {\sc PkANN} to study the response of matter power spectrum to variations in cosmological parameters. {\sc PkANN}'s performance is compared with the $h$-fixed {\sc cosmic emulator} as well. We conclude in Section~\ref{sec:conclusions}. In the Appendix, we detail the formulae used in developing {\sc PkANN}.

\section{Numerical Simulations} \label{sec:SIM}

We run N-body simulations over a range of cosmological parameters with the publicly available adaptive mesh refinement (AMR), grid-based hybrid (hydro+gravity) code {\sc enzo}\footnote{http://lca.ucsd.edu/projects/enzo} (\citealt{NorBryHarBorReySheWag07, OShea10}). All our simulations are hydro+gravity and run in unigrid (AMR switched off) mode. For the hydrodynamical simulations, we include radiative cooling of baryons using an analytical approximation \citep{Sarazin87} for a fully ionized gas with a metallicity of $0.5$ $M_{\sun}$. The cooling approximation is valid over the temperature range from $10^4-10^9$ K. Below $10^4$ K, the cooling rate is effectively zero. We do not account for metal-line cooling, supernova (SN) feedback or active galactic nucleus (AGN) feedback. The parameters we consider are ${\bf I}\equiv(\Omega_{\rm m} h^2, \Omega_{\rm b} h^2, n_s, w, \sigma_8, \sum m_\nu)$, where $h, \Omega_{\rm m}, \Omega_{\rm b}, n_s, w, \sigma_8$ and $\sum m_\nu$ are the present-day normalized Hubble parameter in units of 100 \kmSecMpc, the present-day matter and baryonic normalized energy densities, the primordial spectral index, the constant equation of state parameter for dark energy, the amplitude of fluctuation on an 8$\,h^{-1}$ Mpc scale and the total neutrino mass, respectively. The limits (see Table~\ref{tab:train_set_priors}) on this six-dimensional parameter space includes the WMAP 7-year+BAO+$H_0$ \citep{Komatsu11} constraints.

For details on generating the initial conditions for simulations, and treating massive neutrinos, refer Paper I. Our N-body simulations do not explicitly account for the presence of neutrino perturbations and implement neutrinos only through its effects on the background evolution. Specifically, we modified the cosmological routines of the {\sc enzo} code to include the effects of massive neutrinos on the homogeneous Hubble expansion $h(a)$ (for details, see Paper II) and the linear growth factor. Our modifications to the growth factor neglect any scale-dependence in the presence of massive neutrinos. We sample $70\ (\sum m_\nu=0) + 130\ (\sum m_\nu \ne 0)=200$ (training set), $18+32=50$ (validation set) and $150+180=330$ (testing set) cosmologies from the parameter space (see Table~\ref{tab:train_set_priors}) using an improved Latin hypercube technique (for details, see Paper II). The training set guides the neural network training, the validation set prevents the ANN from overfitting to the training set, and the testing set is used to evaluate the performance of the trained network. The testing set has no effect on training and provides an independent measure of network performance. For each cosmology ${\bf I}\equiv(\Omega_{\rm m} h^2, \Omega_{\rm b} h^2, n_s, w, \sigma_8, \sum m_\nu)$, we compute the Hubble parameter $h$ using the WMAP 7-year+BAO constraint on the acoustic scale ${\pi d_{ls}}/{r_s}=302.54$, where $d_{ls}$ is the distance to the surface of last scattering and $r_s$ is the comoving size of the sound horizon at the redshift of last scattering. The procedure to compute $h$ is outlined in Paper II. This $h$ value, together with the chosen $\Omega_{\rm m} h^2$ and $\Omega_{\rm b} h^2$, is used to derive $\Omega_{\rm m}$ and $\Omega_{\rm b}$. The present-day normalized energy density of dark energy is fixed as $\Omega_{\rm de}=1-\Omega_{\rm m}$. Starting at redshift $z=99$, all simulations are run in a comoving box of length 200 \hmpc, with $256^3$ cold dark matter (CDM) particles evolved on a $512^3$ grid. We take 111 snapshots of the CDM and baryon positions between redshifts $z=2$ and $z=0$; specifically 100 snapshots ($\Delta z=0.01$ apart) between $0 \leq z \leq 0.99$, and 11 snapshots ($\Delta z=0.1$ apart) between $1 \leq z \leq 2$.

\begin{figure}
     \includegraphics[width= \columnwidth]{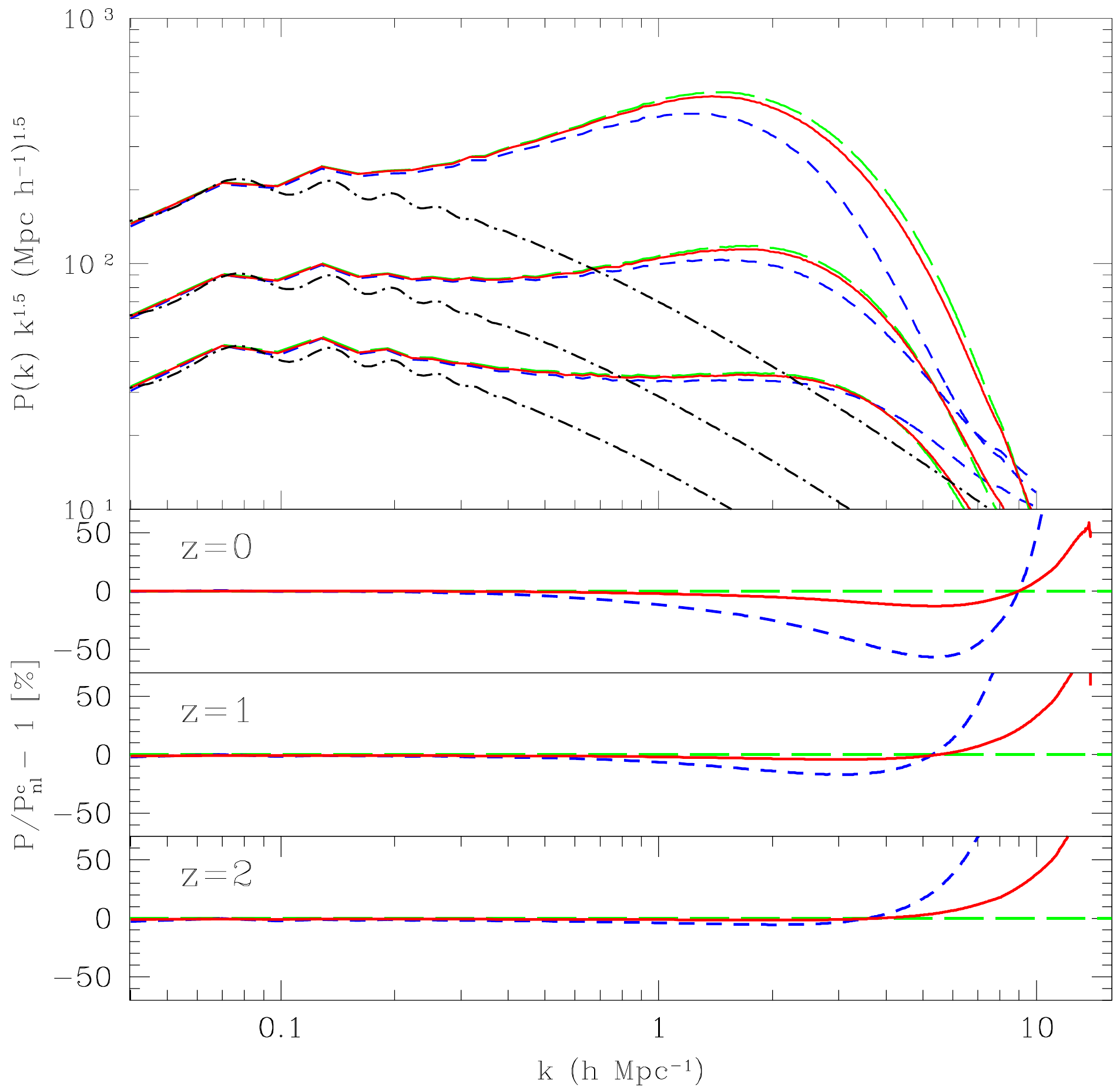}
        \caption{\small{{\it Top panel:} Matter power spectrum evaluated at redshifts $z=0,1,2$ (top to bottom sets, respectively) for the cosmological model ${\bf I}\equiv(0.1196, 0.0232, 0.992, -0.72, 0.8587, 0)$ with $h=0.6496$. At each redshift, the various lines are the non-linear spectra computed using hydro+gravity simulations: (i) $P_{\mathrm{nl}}^{\rm c}$ (long-dashed), (ii) $P_{\mathrm{nl}}^{\rm b}$ (short-dashed) and (iii) $P_{\mathrm{nl}}$ (solid). The linear matter power spectrum is shown by dot-dashed line. $P_{\mathrm{nl}}$ is constructed using $P_{\mathrm{nl}}^{\rm c}$ and $P_{\mathrm{nl}}^{\rm b}$, as discussed in the text (see Eqs~\ref{eq:Pnl} and \ref{eq:Pcb}). {\it Lower panels:} The ratio of the non-linear spectra ($P_{\mathrm{nl}}^{\rm c}, P_{\mathrm{nl}}^{\rm b}$ and $P_{\mathrm{nl}}$) to the CDM spectrum $P_{\mathrm{nl}}^{\rm c}$.
        }}
    \label{fig:cdm_gas}
\end{figure}

Using a Cloud-in-Cell (CIC) interpolation scheme, we transform the CDM and baryon positions into their respective mass density fields. The densities are Fast Fourier Transformed to obtain the CDM and baryon non-linear power spectra, namely $P_{\mathrm{nl}}^{\rm c}$ and $P_{\mathrm{nl}}^{\rm b}$, respectively. Together with the neutrino linear spectrum $P_{\mathrm{lin}}^\nu$, and the weights $f^{\rm i}\equiv\Omega_{\rm i} / \Omega_{\rm m}$, the non-linear matter power spectrum $P_{\mathrm{nl}}$ is then calculated (for details, see Paper I) as
\be
 \label{eq:Pnl}
P_{\mathrm{nl}}(k)=\left[
	(f^{\rm c}+f^{\rm b})\sqrt{P_{\mathrm{nl}}^{\rm cb}(k)}+f^\nu\sqrt{P^\nu_{\mathrm{lin}}(k)}
	\right]^2,
\ee
where,
\be
 \label{eq:Pcb}
P_{\mathrm{nl}}^{\rm cb}(k)=(f^{\rm c}+f^{\rm b})^{-2}\left[f^{\rm c}\sqrt{P^{\rm c}_{\mathrm{nl}}(k)} + f^{\rm b}\sqrt{P^{\rm b}_{\mathrm{nl}}(k)}\right]^2.
\ee
The subscripts `lin' and `nl'  indicate quantities in the linear and non-linear regimes, respectively. Throughout our analyses, we work with flat cosmological models: $\Omega_{\rm m}(=\Omega_{\rm b}+\Omega_{\rm c}+\Omega_\nu)+\Omega_{\rm de}=1$, where $\Omega_{\rm c}$ and $\Omega_\nu$ are the present-day normalized energy densities of CDM and neutrino, respectively. To suppress statistical scatter in the matter power spectrum, we average the power spectra for 11 realizations per cosmology. In Fig.~\ref{fig:cdm_gas}, we show $P_{\mathrm{nl}}^{\rm c}, P_{\mathrm{nl}}^{\rm b}$ and $P_{\mathrm{nl}}$ spectra (long-dashed, short-dashed and solid lines, respectively) for one of the cosmological models ${\bf I}\equiv(0.1196, 0.0232, 0.992, -0.72, 0.8587, 0)$ with $h=0.6496$. The linear matter power spectrum is shown by dot-dashed line. At $k=1\,h \textrm{Mpc}^{-1}$, baryons suppress the CDM spectrum at $1-2$ per cent level. At low redshifts ($z\ltwid2$), as the gas component cools and condenses, it collapses to the center of CDM halos, thereby enhancing the gas power spectrum above the CDM spectrum on smaller scales ($k\gtwid10\,h \textrm{Mpc}^{-1}$). This is consistent with previous studies (\citealt*{RudZenKra2008}, \citealt{CasMacBonSti11}) that investigated the effect of baryonic physics on the matter power spectrum through simulations including gas cooling, star formation and SN feedback. We note that although all our simulations in this work are hydro+gravity, on large scales ($k\ltwid1\,h \textrm{Mpc}^{-1}$) the matter power spectrum is minimally affected by baryonic dynamics and one can rely on gravity-only simulations.

We use the one-Loop standard PT as implemented by \cite{Saito08} for estimating the matter power spectrum up to $k\leq0.085\,h \textrm{Mpc}^{-1}$ and stitch it with the non-linear power spectrum from numerical simulations. Finally, the stitched spectrum is sampled at 50 $k$-values between $0.006\,h \textrm{Mpc}^{-1} \leq k \leq 1\,h \textrm{Mpc}^{-1}$. The stitched-and-sampled non-linear power spectrum is used as $P_{\mathrm{nl}}(k,z)$ for ANN training. This stitch-and-sample procedure is repeated for each cosmology ${\bf I}$ in the training set to complete the training set $P_{\mathrm{nl}}(k,z|{\bf I})$.

\begin{table*}
\caption{\small{Parameter space used in generating the ANN training and validation sets. The last column shows the corresponding WMAP 7-year+BAO+$H_0$ constraints at 68 per cent CL. Inside parentheses is the range for the ANN testing set. The range of the parameters for the testing set is designed to avoid the boundaries of the parameter space. Neutrino mass being physically bound ($\sum m_{\nu} \gtrsim 0$), the lower bound on neutrino mass is set at zero.}}
\begin{tabular}{cccc} \hline
 \multicolumn{1}{c}{Cosmological parameters}& \multicolumn{1}{c}{Lower value} &  \multicolumn{1}{c}{Upper value} &  \multicolumn{1}{c}{WMAP 7-year+BAO+$H_0^a$} \\  \hline
 $\Omega_{\rm m}h^2$		& 0.110 (0.120) 	& 0.165 (0.150) 	& 0.1352		$\pm$ 0.0036	\\
 $\Omega_{\rm b}h^2$		& 0.021 (0.022)		& 0.024 (0.023)		& 0.02255		$\pm$ 0.00054	\\
 $n_{\rm s}$				& 0.85 (0.90)		& 1.05 (1.00)		& 0.968		$\pm$ 0.012	\\
 $w$					& -1.35 (-1.15)		& -0.65 (-0.85)		& -1.1		$\pm$ 0.14	\\
 $\sigma_8$				& 0.60 (0.70)		& 0.95 (0.85)		& 0.816		$\pm$ 0.024	\\
 $\sum m_\nu$ (eV)			& 0 (0)			& 1.1	 (0.5)			& $<0.58^b$			\\ \hline
 \multicolumn{1}{l}{{\it Note.} $^a$\cite{Komatsu11}; $^b95\,$ per cent CL for $w=-1.$}
\end{tabular}
\label{tab:train_set_priors}
\end{table*}

\section{Artificial Neural Networks}
\label{sec:ANN}

\begin{figure}
     \includegraphics[width= \columnwidth]{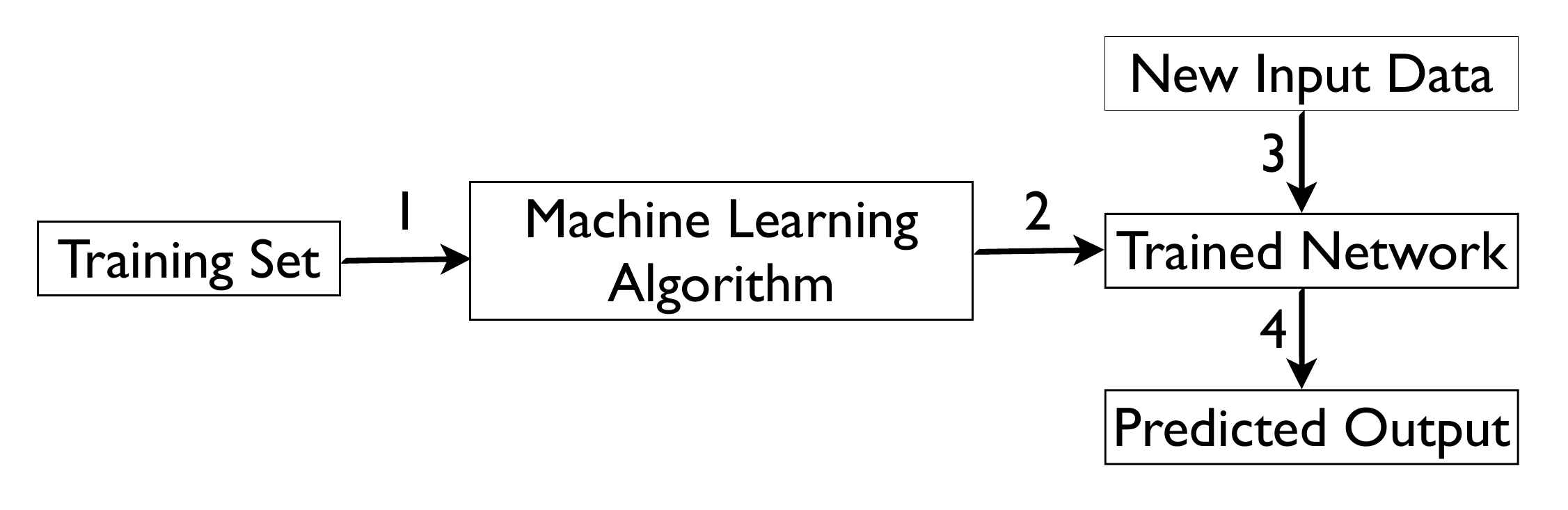}
        \caption{\small{Steps 1 and 2: A machine-learning network learns to parameterize the output, for the input patterns that form the training set. Steps 3 and 4: The trained network is capable of making predictions when presented with input parameter settings. The queried input settings must lie within the parameter ranges of the patterns in the training set.
        }}
    \label{fig:ANN1}
\end{figure}

Fig.~\ref{fig:ANN1} shows a skeleton of a machine-learning network. Using a suitable training set (input parameters for which data is available), the machine-learning algorithm is trained to learn a parameterization. With this parameterization the network is capable of reproducing (as closely as possible) the output, when queried with input parameter settings that are part of the training set. The trained network can now be presented with new settings of the input parameters (for which one does not have any prior data) and by using the same parameterization learnt during the training process, the network makes predictions.

ANN -- a form of machine-learning -- is a collection of {\it nodes} arranged in a series of layers, with each node in a layer connected to all other nodes in adjacent layers. A network's architecture is specified by the number of nodes from input to output as $N_{in}$ : $N_1$ : $N_2$ : ... : $N_n$ : $N_{out}$. That is, a network with an architecture 4 :  9 : 5 : 7 has 4 inputs, two hidden layers with 9 and 5 nodes respectively, and finally 7 outputs. An extra node (called the {\it bias} node) is added to the input layer as well as to each of the hidden layers. The bias nodes are added in order to compensate for the difference between the network's mean prediction and the mean of the outputs of training set patterns (for details, refer \citealt{Bishop95}). Each bias node connects to all the nodes in the next layer. Note that the counts $N_{in}, N_1, N_2, ..., N_n$ do not include the bias nodes. The output layer has no bias node. The total number of connections (also called the {\it weights}) $N_W$ for a generic architecture $N_{in}$ : $N_1$ : $N_2$ : ... : $N_n$ : $N_{out}$ can be calculated using the formula
\be
 \label{eq:Nodes}
N_W=N_{in}\cdot N_1 + \sum_{l=2}^n N_{l-1}\cdot N_l + N_n \cdot N_{out} + \sum_{l=1}^n N_l + N_{out},
\ee
where the summation index $l$ is over the hidden layers only. Throughout this paper, we will use the vector notation ${\bf w}$ to collectively refer to all the network weights.

\begin{figure}
     \includegraphics[width= \columnwidth]{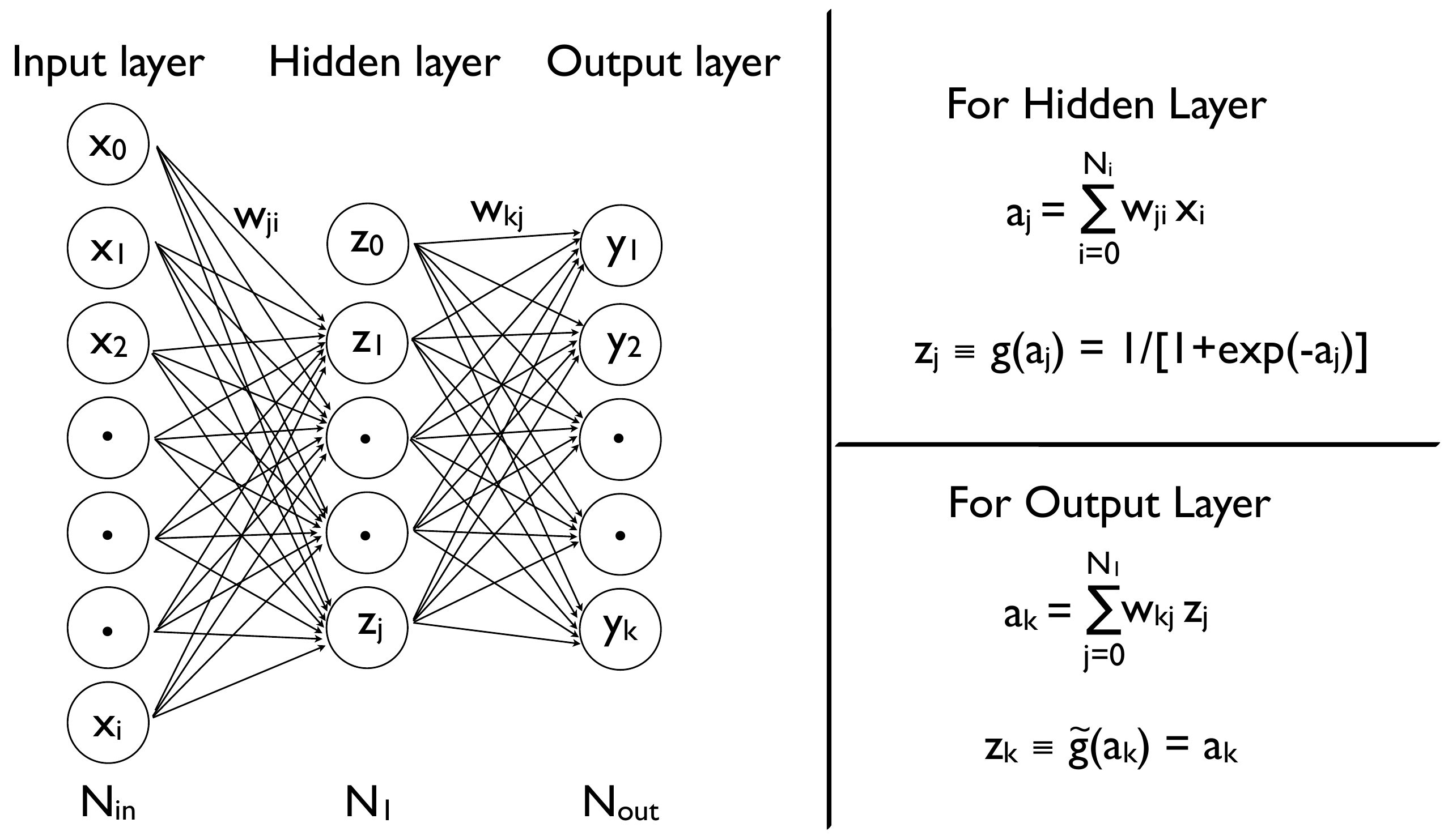}
        \caption{\small{A typical ANN architecture (left-hand panel) with node activation formulae for the hidden and output layers (right-hand panels). There can be more than one hidden layers. Throughout our {\sc PkANN} analysis, we work with a single hidden layer.
        }}
    \label{fig:ANN2}
\end{figure}

In Fig.~\ref{fig:ANN2}, we show a typical ANN architecture (left-hand panel) and the formulae to calculate the node activations (right-hand panels). In the network configuration depicted, there are $N_{in}$ input parameters/features $(x_1,...,x_i)$, a single hidden layer with $N_1$ nodes $(z_1,...,z_j)$, and $N_{out}$ output parameters/features $(y_1,...,y_k)$. The bias nodes in the input and hidden layers are $x_0$ and $z_0$, respectively.

Each node in the $l$th hidden layer is a neuron with an $\it{activation}$, $z_j\equiv g(a_j)$, taking as its argument
\be
\label{eq:aj}
a_j=\displaystyle\sum_{i=0} w_{ji}z_i,
\ee
where the sum is over all nodes $i$ (including the bias node) of the previous layer sending connections to the $j$th node (barring the bias node) of the current layer. Note that for networks with a single hidden layer (as in Fig.~\ref{fig:ANN2}), $z_i$ in Eq.~\ref{eq:aj} would correspond to the input parameters $x_i$. The activation functions are typically taken to be sigmoid functions such as $g(a_j) = 1/[1 +\rm exp(-a_j)]$. Since the range of $g(a_j)$ is from 0 to 1, it allows the output of the neurons to be interpreted as the probability that any specific neuron will `fire' when presented with an input parameters setting. The sigmoid functions impart some degree of non-linearity to the neural network models. A network becomes overly non-linear if the weights ${\bf w}$ deviate significantly from zero. This drives the activation $g(a_j)$ of the nodes to saturation. The number and size of the hidden layers add to the complexity of ANNs. The activation of all bias nodes is permanently set to a value of 1 and during network training the bias parameters (namely, $w_{j0}$ and $w_{k0}$ in Fig.~\ref{fig:ANN2} left-hand panel) are adjusted so as to minimize the difference between the mean prediction for the network and the mean of the outputs of the training set patterns.

The activation $y_k\equiv \tilde{g}(a_k)$ for neurons in the output layer is usually taken to be $a_k$, \ie $\tilde{g}(a_k)=a_k$, with $a_k$ being the weighted sum of all nodes in the final hidden layer,
\be
\label{eq:ak}
a_k=\displaystyle\sum_{j=0} w_{kj}z_j.
\ee
For a particular input vector $(x_1,...,x_i)$, the output vector $(y_1,...,y_k)$ of the network is determined by progressing sequentially through the network layers, from inputs to outputs, calculating the activation of each node.

Adjusting the weights ${\bf w}$ to get the desired mapping is called the {\it training} of the network. For matter power spectrum estimation, we use a training set of N-body simulations with known cosmological parameters:
\be
\label{eq:PARAMS} \nonumber
{\bf I}\equiv(\Omega_{\rm m} h^2, \Omega_{\rm b} h^2, n_s, w, \sigma_8, \sum m_\nu).
\ee

It has been shown (see \citealt{Hornik91,Ito91,Bishop95}) that networks with a single hidden layer are capable of making arbitrarily accurate approximation to a function and its derivatives. As such, for {\sc PkANN}'s architecture, we only consider networks having single-hidden layer with sigmoidal activations and output nodes with linear ($\tilde{g}(a_k)=a_k$) activations, as depicted in Fig.~\ref{fig:ANN2}.

In Appendix~\ref{sec:PkANN}, we develop the {\sc PkANN} cost function $\chi^2_{C}({\bf w})$. Minimizing this cost function with respect to the weights {\bf w} generates a trained ANN that can be used for non-linear matter power spectrum interpolation. To minimize $\chi^2_{C}$({\bf w}) (see Eq.~\ref{eq:cost}) with respect to the weights {\bf w}, we use an iterative quasi-Newton algorithm (Appendix \ref{sec:QNM}) that involves evaluating the first-order derivative (gradient) of the cost function. See Appendix \ref{sec:CFG} for the derivation of the gradient. The quasi-Newton algorithm also involves information about the inverse of the Hessian (second-order derivative) matrix which we approximate using the Broyden--Fletcher--Goldfarb--Shanno (BFGS) method (see Appendix \ref{sec:CFH}. For details, see \citealt{Bishop95}).

Starting with randomly assigned weights ${\bf w}$, their values are re-estimated iteratively, making sure that each iteration proceeds in a direction that lowers the cost function $\chi^2_{C}({\bf w})$. In order to avoid over-fitting to the training set, after each iteration to the weights, Eq.~\ref{eq:cost} is also calculated for what is known in neural network parlance as a validation set. The validation set for our application of neural networks, is a small set of simulations with known ${\bf I}\equiv(\Omega_{\rm m} h^2, \Omega_{\rm b} h^2, n_s, w, \sigma_8, \sum m_\nu)$ and $P_{\mathrm{nl}}(k,z)$. The final weights ${\bf w}_f$ are chosen so as to give the best fit (minimum $\chi^2_{C}({\bf w})$) to the validation set. The network training is considered finished once $\chi^2_{C}({\bf w})$ is minimized with respect to the validation set. The trained network can now be used to predict $P_{\mathrm{nl}}(k,z)$ for new cosmologies. It is important to note that starting with a different (but still random) configuration of weights, may lead to a trained network with a different set of final weights ${\bf w}_f$. As such, we train a number of networks that start with an alternative random configuration of weights. The trained networks are collectively called a {\it committee} of networks and subsequently give rise to better performance than any single ANN in isolation. For the final output, we average over the outputs of the committee members.

\section{Results}
\label{sec:RESULTS}

\subsection{Comparing PkANN against Numerical Simulations}
\label{sec:PkANN_vs_SIM}

In Paper II, we compared {\sc PkANN}'s performance against {\sc halofit} spectra to demonstrate that a suitably trained network is capable of reproducing the {\sc halofit} spectra at sub-per cent accuracy. Here, we repeat the procedure, this time using spectra calculated using N-body simulations. We selected the combination 7 :  $N_{hidden}$ :  50 as our {\sc PkANN} architecture, where $N_{hidden}$ (number of nodes in the hidden layer) was varied from 7 to 98, in steps of 7. The number of inputs were fixed at 7, corresponding to ${\bf I}\equiv(\Omega_{\rm m} h^2, \Omega_{\rm b} h^2, n_s, w, \sigma_8, \sum m_\nu)$ including redshift $z$. As discussed in Section~\ref{sec:SIM}, we use the {\sc camb} code to calculate the CDM, baryon and neutrino transfer functions. The initial conditions for CDM particles and baryons are then generated from their transfer functions using {\sc enzo}. The non-linear matter power spectrum $P_{\mathrm{nl}}(k)$ is constructed using Eqs~\ref{eq:Pnl} and \ref{eq:Pcb}.

\begin{figure*}
     \includegraphics[width= 6in, height=6in]{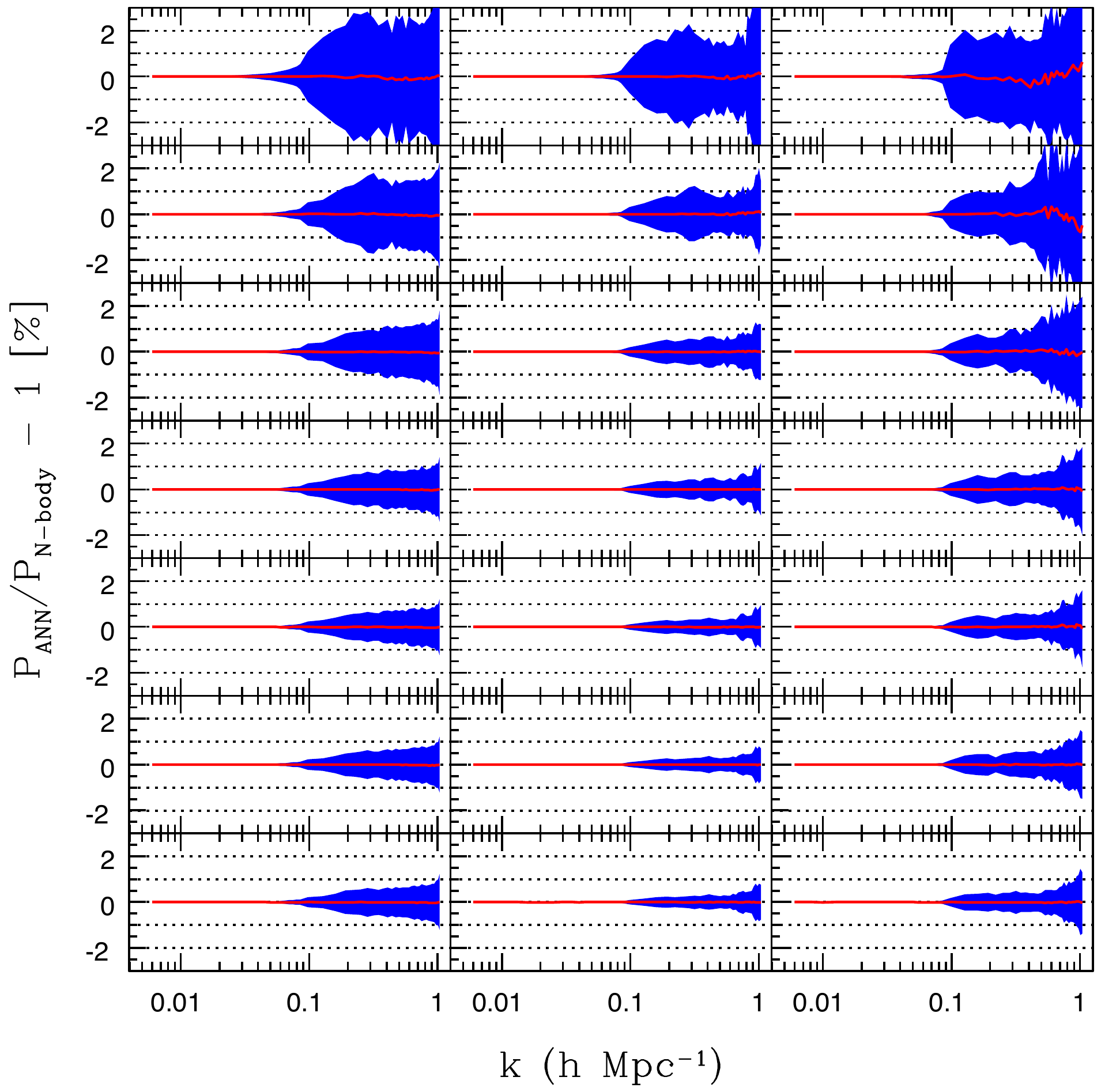}
        \caption{\small{Percentage error at redshift $z=0$ (left-hand panel), $z=1$ (middle panel) and $z=2$ (right-hand panel) between the predicted non-linear power spectrum (using {\sc PkANN}) and the true underlying spectrum (using N-body simulations) for $200$ training set cosmologies. The shaded region contains the middle 99.73\% ($3\sigma$) of the residuals. The rows (from top to bottom) correspond to $N_{hidden}=14-98$ in increments of 14. The mean error over all 200 cosmologies is shown by a solid line -- an indicator of any bias in the ANN training scheme.
    }}
    \label{fig:fit_error_training}
\end{figure*}

As in Paper II, we do not sample the redshift in the Latin hypercube but instead evaluate $P_{\mathrm{nl}}(k,z)$ at 111 redshifts between $z=0$ and $z=2$ from numerical simulations, using Eqs~\ref{eq:Pnl} and \ref{eq:Pcb}. As we discussed in Section~\ref{sec:SIM}, we extend the range of our spectra to $k=0.006\,h \textrm{Mpc}^{-1}$ by using the one-loop standard PT \cite{Saito08}. We estimate the matter power spectrum up to $k\leq0.085\,h \textrm{Mpc}^{-1}$ using the one-loop standard PT and stitch it with $P_{\mathrm{nl}}(k,z)$. The stitched spectrum is then sampled at 50 $k$-modes between $0.006\,h \textrm{Mpc}^{-1} \leq k \leq 1\,h \textrm{Mpc}^{-1}$. Since our training and validation sets have ($130+70$) and ($32+18$) cosmologies, respectively (see Paper II), we calculated $P_{\mathrm{nl}}(k,z)$ for each cosmology, at 111 redshifts. These $P_{\mathrm{nl}}(k,z)$ are scaled by their respective linear spectra $P_{\mathrm{lin}}(k,z)$ (see Eq \ref{eq:chisq}), before being fed to the neural network. Thus, the overall size $N_T$ of the training set that we train our ANN with is $N_T=200\times111=22,200$. Likewise, we have $50\times111=5,550$ patterns in the validation set. For each $N_{hidden}$ setting, we trained a committee of 16 ANNs. The weights $\bf w$ for each ANN were randomly initialized (the random configuration being different for each ANN). The weights are allowed to evolve until $\chi^2_{C}({\bf w})$ (see Eq.~\ref{eq:cost}) is minimized with respect to the cosmologies in the validation set.

In Fig.~\ref{fig:fit_error_training}, we show the percentage error in the ANN predictions with respect to the N-body results when presented with the 200 cosmologies in the training set. We average the $P_{\mathrm{nl}}^{\rm ANN}(k,z)$ predictions over the 16 ANN committee members. The rows correspond to $N_{hidden}=14-98$ (from top to bottom) in increments of 14. The columns (from left to right) correspond to $z=0,1,2$. The mean error over all 200 cosmologies in the training set is shown by a solid line in each panel, to get an idea about any systematics in our ANN training scheme. With $N_{hidden}=70$ and higher, the ANN predictions are within $\pm 1$ per cent of the N-body power spectra for $k\leq0.9\,h \textrm{Mpc}^{-1}$, after which the performance degrades marginally to $\pm 1.5$ per cent. The worst-performing cosmologies correspond to the parameter settings with at least four of the six cosmological parameters at their boundary values.

\begin{figure*}
      \includegraphics[scale=0.4]{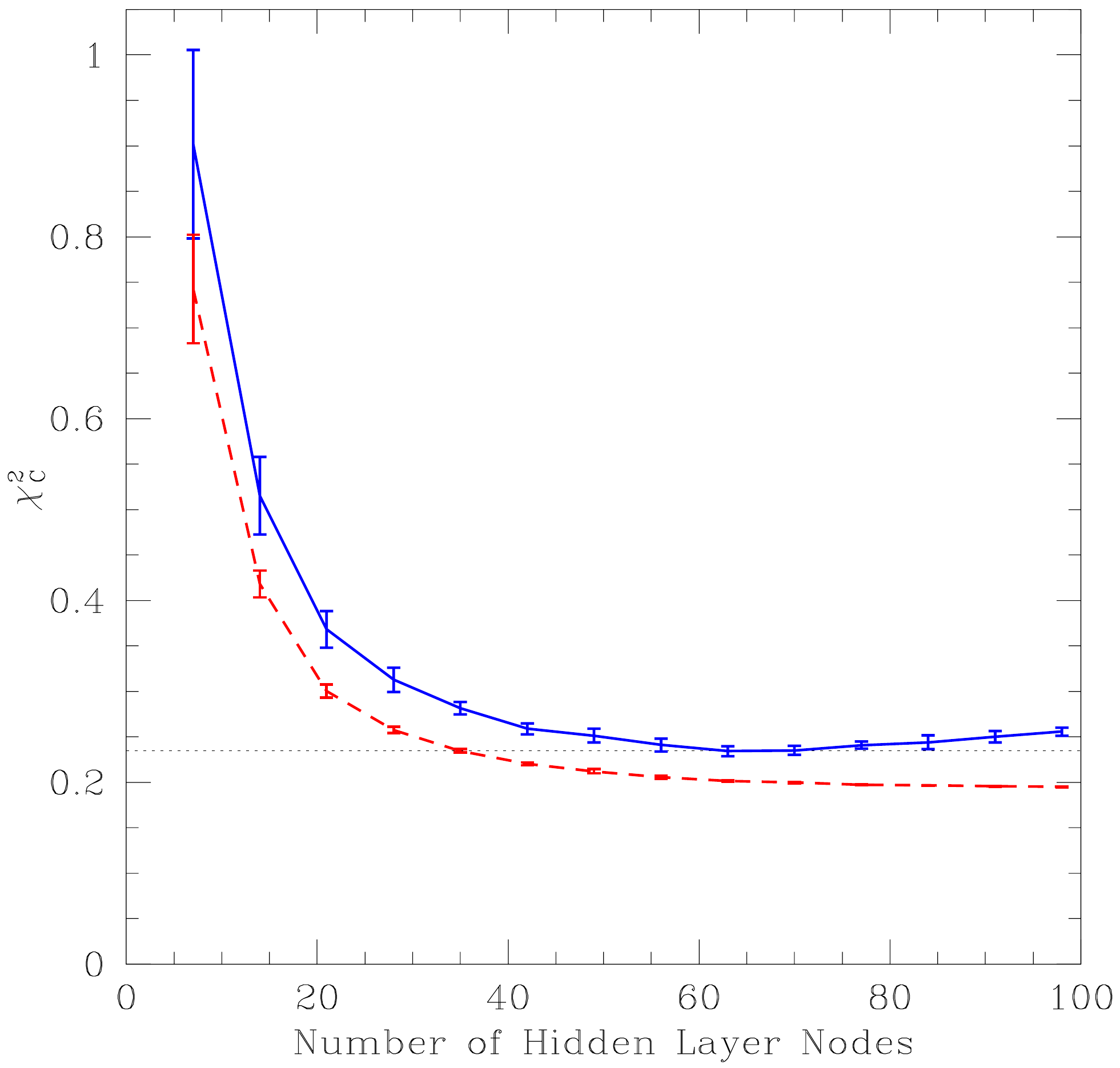}
        \caption{\small{The residual error $\chi^2_{C}({\bf w})$ (see Eq.~\ref{eq:cost}) evaluated as a function of the number of nodes in the hidden layer, $N_{hidden}$. The error is a monotonically decreasing function for the training set (dashed line) while for the validation set (solid line), it starts increasing beyond $N_{hidden}=70$ indicating that the generalizing ability of the neural network is best with $N_{hidden}=70$. The error bars correspond to the spread in $\chi^2_{C}({\bf w})$ for the 16 ANN committee members.
        }}
    \label{fig:regularization_enzo}
\end{figure*}

\begin{figure*}
     \includegraphics[width= 6in, height=6in]{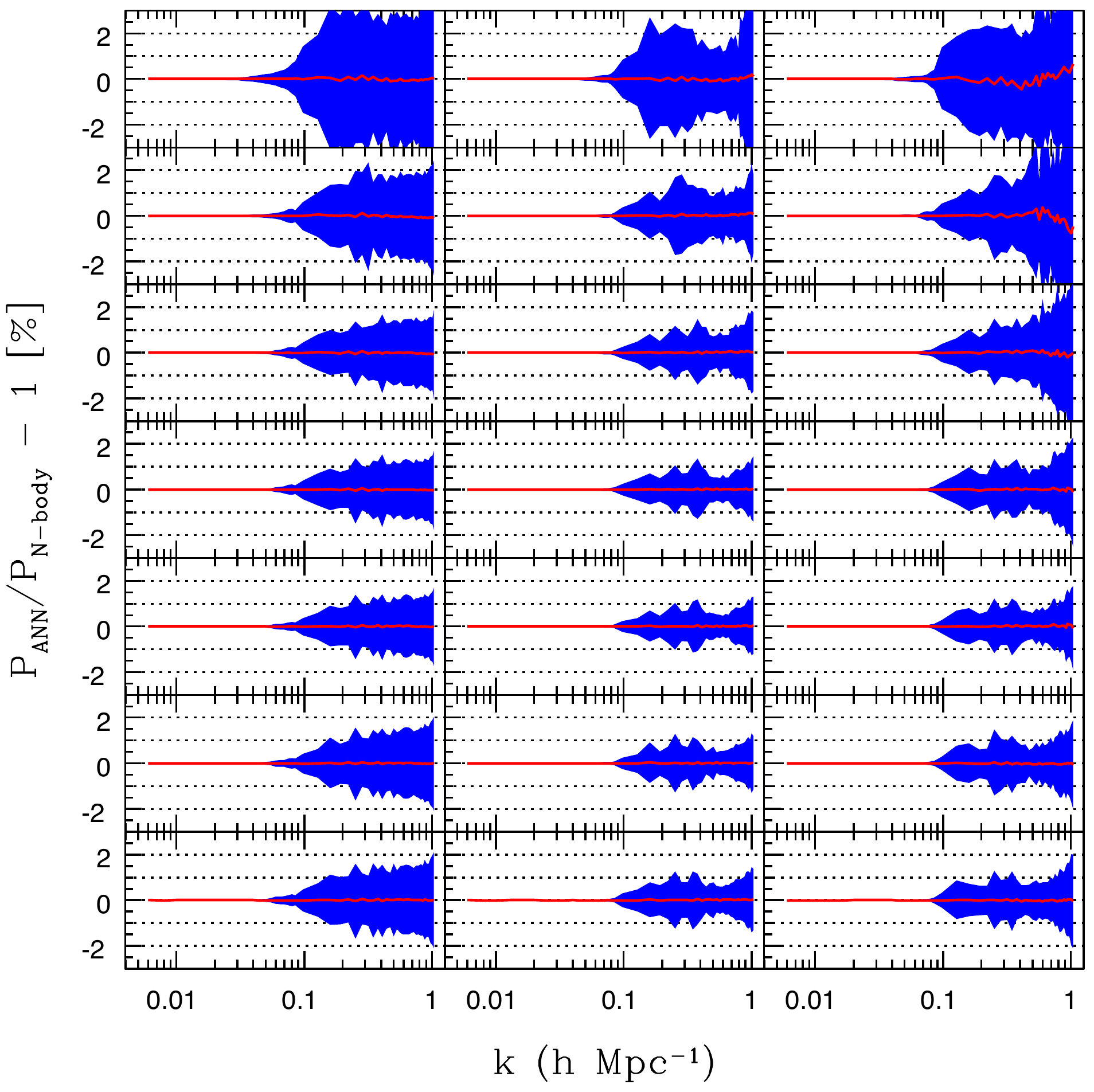}
        \caption{\small{Similar to Fig.~\ref{fig:fit_error_training}, using 50 validation set cosmologies.
    }}
    \label{fig:fit_error_validation}
\end{figure*}

Increasing the number of nodes in the hidden layer increases the flexibility of a neural network. An increasingly complex network can make extremely accurate predictions on the training set. This is evident from Fig.~\ref{fig:fit_error_training}, where the prediction over the training set becomes progressively better (from top to bottom) with increasing $N_{hidden}$ units. However, such complex networks can adversely affect their generalizing ability when presented with a new dataset. The validation set helps in controlling the complexity of a network, as we discussed earlier in Section~\ref{sec:ANN}. In Fig.~\ref{fig:regularization_enzo}, we show the residual cost function $\chi^2_{C}({\bf w})$ (see Eq.~\ref{eq:cost}) evaluated as a function of the number of nodes in the hidden layer, $N_{hidden}$. The residual error is a monotonically decreasing function for the training set (dashed line) while for the validation set (solid line), it increases beyond $N_{hidden}=70$. The performance of the trained ANNs as a function of $N_{hidden}$ units, over the cosmologies in the validation set, is shown in Fig.~\ref{fig:fit_error_validation}.  Increasing $N_{hidden}$ beyond 70 increases the error marginally, indicating that $N_{hidden}=70$ saturates the generalizing ability of our network. 

\begin{figure*}
     \includegraphics[width= 6in, height=6in]{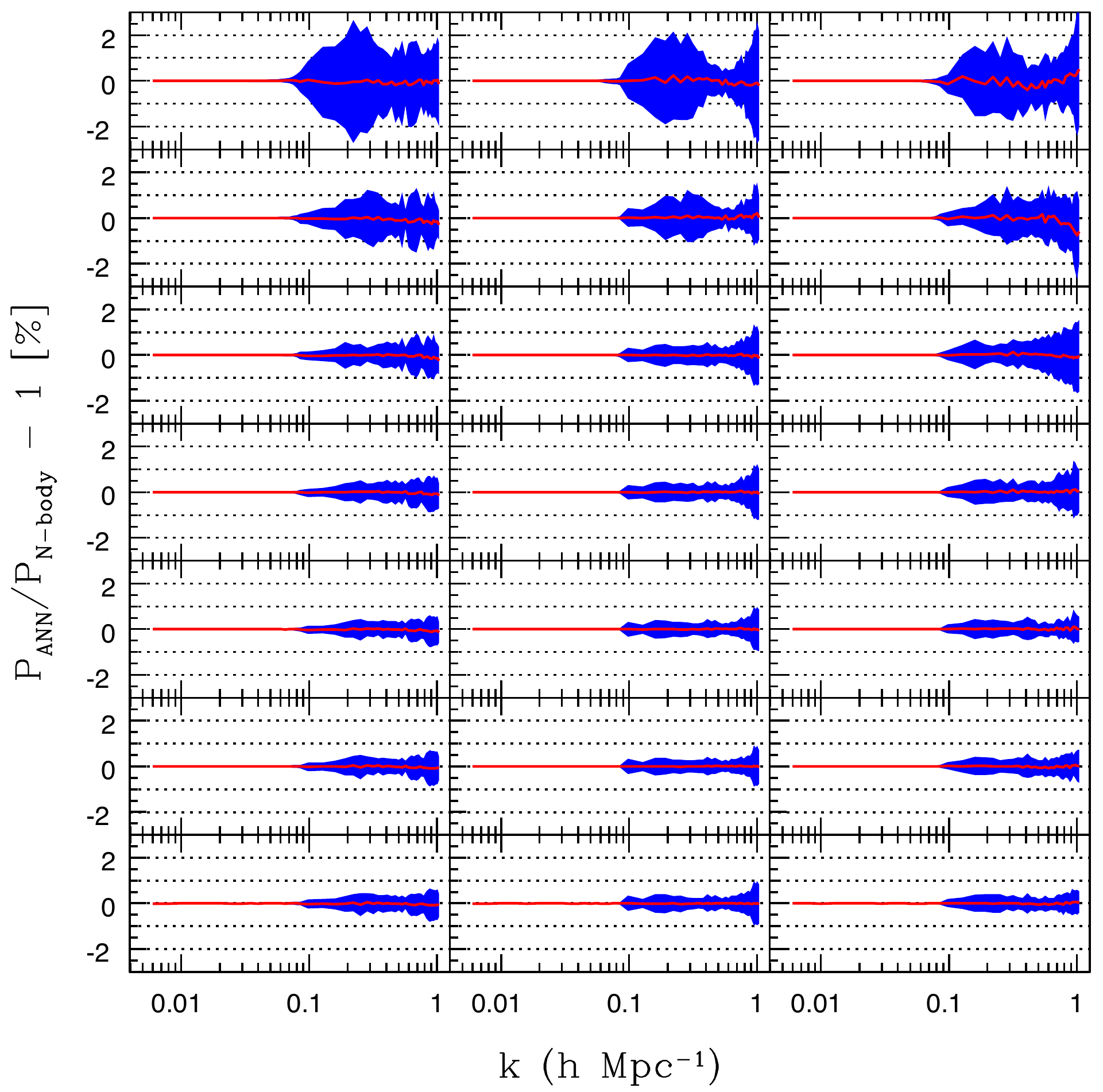}
        \caption{\small{Similar to Fig.~\ref{fig:fit_error_training}, using 330 testing set cosmologies.
    }}
    \label{fig:fit_error_testing}
\end{figure*}

The performance of the trained ANNs for cosmological models in the testing set, is shown in Fig.~\ref{fig:fit_error_testing}. Increasing $N_{hidden}$ beyond 70 does not contribute to a significant error reduction on the testing set, confirming our assessment that $N_{hidden}=70$ saturates the generalizing ability of the network. With $N_{hidden}=70$, the ANN prediction for {\it every} cosmology, at {\it all} redshifts $z\leq2$, is within $\pm 0.5$ per cent of the N-body power spectra up to $k\leq0.9\,h \textrm{Mpc}^{-1}$. The {\sc PkANN} performs exceedingly well within the boundaries of the restricted parameter space.

\begin{figure*}
     \includegraphics[width= 6in, height=6in]{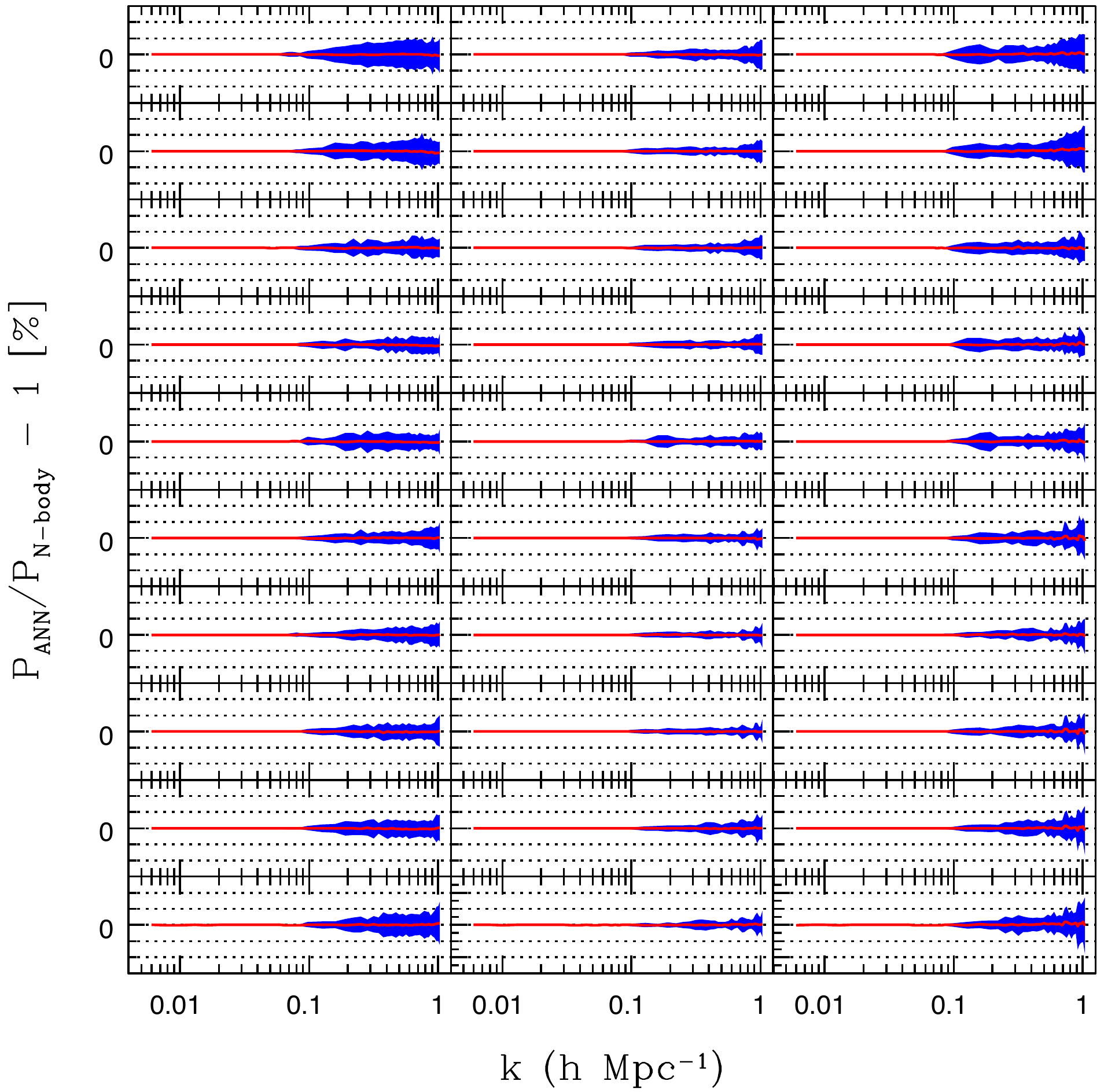}
        \caption{\small{The 200 cosmologies of the training set are binned in 10 equal intervals between $\Omega_{\rm m} h^2=0.11-0.165$ (from top to bottom, in increasing order). The columns are redshift $z=0,1,2$ (from left to right, respectively). $N_{hidden}=70$ for all panels. For each bin, {\sc PkANN}'s predictions are compared to the N-body power spectra and the residual errors ($3\sigma$ CL) are plotted. Closer to the middle of the range $\Omega_{\rm m} h^2=0.11-0.165$ (middle rows), the prediction errors get smaller. Even near the edges (outer rows), the errors are within $\pm 1$ per cent of the N-body power spectra for $k\leq0.9\,h \textrm{Mpc}^{-1}$.
    }}
    \label{fig:ps_fit_enzo_bin_1}
\end{figure*}

Next, we assess the accuracy of the {\sc PkANN} network across the range for each of the six parameters, namely, $\Omega_{\rm m} h^2, \Omega_{\rm b} h^2, n_s, w, \sigma_8$ and $\sum m_\nu$. We vary each parameter between its minimum and maximum values and bin the 200 cosmologies of the training set in 10 intervals across the parameter range. We calculate the prediction error for each bin. We repeat this for all six parameters and show the results for the $\Omega_{\rm m} h^2$ case in Fig.~\ref{fig:ps_fit_enzo_bin_1}. The rows correspond to the 10 linearly spaced bins between $\Omega_{\rm m} h^2=0.11-0.165$. The columns are redshift $z=0$ (left-hand panel), $z=1$ (middle panel) and $z=2$ (right-hand panel). As discussed above, we fix $N_{hidden}=70$. As expected, {\sc PkANN}'s performance degrades near the edges of the range $\Omega_{\rm m} h^2=0.11-0.165$ (compare the middle rows against the outer rows). Overall, the prediction errors remain within $\pm 1$ per cent of the N-body power spectra for $k\leq0.9\,h \textrm{Mpc}^{-1}$. Results with the other five parameters are similar to Fig.~\ref{fig:ps_fit_enzo_bin_1}. We summarize the prediction errors for all six parameters in Table~\ref{tab:Summary_All}.

\begin{table*}
\caption{\small{Performance of the {\sc PkANN} network as a function of the range of the six parameters, namely, $\Omega_{\rm m} h^2, \Omega_{\rm b} h^2, n_s, w, \sigma_8$ and $\sum m_\nu$. Each parameter range is sub-divided into 10 equal intervals and the training set cosmologies are binned accordingly. The $3\sigma$ bounds on the {\sc PkANN} prediction errors (in per cent) are mentioned for each bin, at redshifts $z=0,1$ and 2. The $\Omega_{\rm m} h^2$ case is shown in Fig.~\ref{fig:ps_fit_enzo_bin_1}.
         }}
\begin{tabular}{ccccccccccccccccccc} \hline
 \multicolumn{1}{c}{Bins}	& \multicolumn{3}{c}{$\Omega_{\rm m} h^2$} &  \multicolumn{3}{c}{$\Omega_{\rm b} h^2$} &  \multicolumn{3}{c}{$n_s$} &  \multicolumn{3}{c}{$w$} &  \multicolumn{3}{c}{$\sigma_8$} &  \multicolumn{3}{c}{$\sum m_\nu$} \\
 \multicolumn{1}{c}{}	& \multicolumn{1}{c}{z=0} 	& \multicolumn{1}{c}{z=1} 	& \multicolumn{1}{c}{z=2} & \multicolumn{1}{c}{z=0} & \multicolumn{1}{c}{z=1} & \multicolumn{1}{c}{z=2} & \multicolumn{1}{c}{z=0} & \multicolumn{1}{c}{z=1} & \multicolumn{1}{c}{z=2} & \multicolumn{1}{c}{z=0} & \multicolumn{1}{c}{z=1} & \multicolumn{1}{c}{z=2} & \multicolumn{1}{c}{z=0} & \multicolumn{1}{c}{z=1} & \multicolumn{1}{c}{z=2} & \multicolumn{1}{c}{z=0} & \multicolumn{1}{c}{z=1} & \multicolumn{1}{c}{z=2} \\  \hline
 1		& 1.0		& 1.0		& 1.0		& 0.9		& 0.7		& 1.1		& 1.0		& 0.9		& 1.2		& 0.9		& 0.9		& 0.9		& 0.8		& 0.5		& 1.0		& 0.8		& 0.8		& 1.1		\\
 2		& 1.0		& 0.8		& 1.1		& 0.6		& 0.7		& 1.1		& 0.9		& 0.8		& 0.9		& 0.8		& 0.8		& 0.9		& 0.8		& 0.5		& 1.0		& 0.8		& 0.7		& 1.0		\\
 3		& 0.9		& 0.9		& 0.9		& 0.9		& 0.7		& 1.0		& 0.9		& 0.8		& 0.9		& 0.7		& 0.7		& 0.9		& 0.6		& 0.5		& 1.0		& 0.5		& 0.4		& 1.0		\\
 4		& 0.7		& 0.8		& 0.9		& 0.8		& 0.6		& 0.9		& 0.8		& 0.6		& 0.9		& 0.6		& 0.7		& 0.9		& 0.6		& 0.5		& 1.0		& 0.5		& 0.6		& 0.9		\\
 5		& 0.7		& 0.5		& 0.9		& 0.7		& 0.6		& 0.9		& 0.6		& 0.5		& 1.0		& 0.6		& 0.6		& 0.9		& 0.5		& 0.5		& 0.9		& 0.7		& 0.5		& 0.9		\\
 6		& 0.9		& 0.5		& 1.0		& 0.4		& 0.5		& 0.8		& 0.7		& 0.5		& 1.0		& 0.7		& 0.5		& 0.9		& 0.4		& 0.5		& 0.8		& 0.8		& 0.7		& 1.0		\\
 7		& 0.8		& 0.6		& 0.9		& 0.7		& 0.6		& 1.0		& 0.7		& 0.8		& 1.0		& 0.7		& 0.7		& 0.9		& 0.5		& 0.4		& 0.9		& 0.9		& 0.5		& 0.9		\\
 8		& 1.0		& 0.6		& 0.9		& 0.8		& 0.6		& 0.9		& 0.6		& 0.7		& 1.0		& 0.8		& 0.5		& 0.8		& 0.6		& 0.4		& 1.0		& 0.9		& 0.8		& 0.9		\\
 9		& 0.9		& 0.8		& 1.0		& 0.8		& 0.6		& 1.0		& 0.8		& 0.8		& 1.0		& 0.9		& 0.7		& 0.9		& 0.8		& 0.5		& 0.9		& 0.9		& 0.8		& 0.9		\\
 10		& 1.0		& 0.8		& 1.1		& 0.9		& 0.7		& 1.1		& 0.8		& 0.8		& 1.0		& 0.9		& 0.6		& 0.9		& 1.0		& 0.7		& 0.9		& 0.8		& 0.9		& 1.2		\\ \hline
\end{tabular}
\label{tab:Summary_All}
\end{table*}

\subsection{PkANN Error Estimates}
\label{sec:ERRORS}

Our ANN framework successfully recreates the input power spectrum at sub-percent level up to $k\leq0.9\,h \textrm{Mpc}^{-1}$, and the overall accuracy of the {\sc PkANN} interpolator is set by the force resolution and statistical variance from our N-body simulations.  Running {\sc enzo} in a 200 \hmpc\ box with $512^3$ unigrid results in a matter power spectrum that is progressively suppressed from 1 per cent level at $k=0.5\,h \textrm{Mpc}^{-1}$ to 5 per cent level at $k=0.9\,h \textrm{Mpc}^{-1}$, when compared to spectrum calculated from high-resolution runs. Limited computing resources prohibited us from running higher resolution simulations. Since {\sc PkANN} is built using conservative simulation settings described above, we expect all {\sc PkANN} predictions to be suppressed at 1-5 per cent level between $k=0.5-0.9\,h \textrm{Mpc}^{-1}$.

We follow the approach outlined in \cite*{Jeong09} (see their Appendix A) to roughly estimate the statistical error on our non-linear power spectrum from numerical simulations. A simulation box of length 200 \hmpc\ corresponds to a fundamental wavenumber of $\delta k = 2\pi/200= 0.0314\,h \textrm{Mpc}^{-1}$. The number of independent $k$-modes available in a spherical shell at $k=0.1\,h \textrm{Mpc}^{-1}$ is $N_k=2\pi(k/\delta k)^2\approx64$. With our 11 realizations per cosmology, this gives a relative error of $\sigma_{P(k)/P(k)}=1/\sqrt{11N_k}\approx4\%$ at $k=0.1\,h \textrm{Mpc}^{-1}$. Higher $k$-modes are sampled more frequently and the corresponding sampling errors become progressively smaller, to $\sim 0.4\%$ at $k=0.9\,h \textrm{Mpc}^{-1}$.

As mentioned earlier, we match the matter power spectra from one-Loop standard PT with numerical simulations at $k=0.085\,h \textrm{Mpc}^{-1}$. \cite{Heitmann10} (their Fig. 6) showed that small simulation volumes fail to capture linear evolution on the largest scales probed by the simulation box as well as miss the onset of non-linearity, resulting in the suppression of the matter power spectrum at $\sim2-3$ per cent level. As such, for a simulation box of length 200 \hmpc, we expect our spectra amplitudes to be in error at $\sim3$ per cent level around $k\approx1\,h \textrm{Mpc}^{-1}$.

{\sc PkANN} can be used for spatially flat cosmological models with three species of degenerate massive neutrinos up to $\sum m_\nu=1.1$ eV. Since our implementation of neutrinos in numerical simulations does not take into account the non-linear evolution of neutrino perturbations, this is expected to introduce errors in the estimated matter power spectrum. In Paper I, we discussed the expected errors by comparing our results with \cite{Brandbyge08} and \cite*{Brandbyge09}. At redshift $z=0$, our neutrino spectra for $\sum m_\nu$ up to $0.1,0.475$ and $0.95$ eV are expected to be in error by $\ltwid 0.1,4$ and $10$ per cent, respectively. The respective errors at $z=1$ and $z=2$ are $\ltwid 0.1,3,6$ and $\ltwid 0.1,3,5$ per cent. These error estimates are large for $\sum m_\nu>0.475$ eV; however, it is important to note that the current constraints on the total neutrino mass are around 0.3 eV. Using photometric redshifts measured from Sloan Digital Sky Survey III Data Release Eight (SDSS DR8, \citealt{SDSS11}), \cite{dePut12} obtained constraints of $\sum m_\nu<0.26-0.36$ eV. Using BAO and CMB data, the Planck survey \citep{Planck13} finds an upper limit of 0.23 eV. Using numerical simulations, \cite*{WagVerJim12} studied the effect of neutrinos on the non-linear matter power spectrum for $\sum m_\nu\leq0.3$ eV and found very similar results as ours in Paper I. For such low neutrino masses ($\sum m_\nu\leq0.3$ eV), \cite*{Brandbyge09} (their Fig. 1) show that at $z=0$ non-linear neutrino corrections are at 0.3 per cent level, and negligible at higher redshifts. Overall, for $\sum m_\nu\leq1.2$ eV, corrections are at 1.5 per cent level for $z\geq1$.

To summarize, across all cosmological models (see Table~\ref{tab:train_set_priors}) with $\sum m_\nu<0.5$ eV, the {\sc PkANN} interpolator is expected to be accurate at 5 per cent level for all redshifts $z\leq2$. For models with $\sum m_\nu>0.5$ eV, the spectra predictions are expected to be accurate at 5 per cent level only for $z>1$ and degrade to $\sim10$ per cent for $z\leq1$.

\subsection{Exploring Cosmological Parameter Space with PkANN }
\label{sec:EXPLORE}

\begin{figure*}
     \includegraphics[scale=0.8]{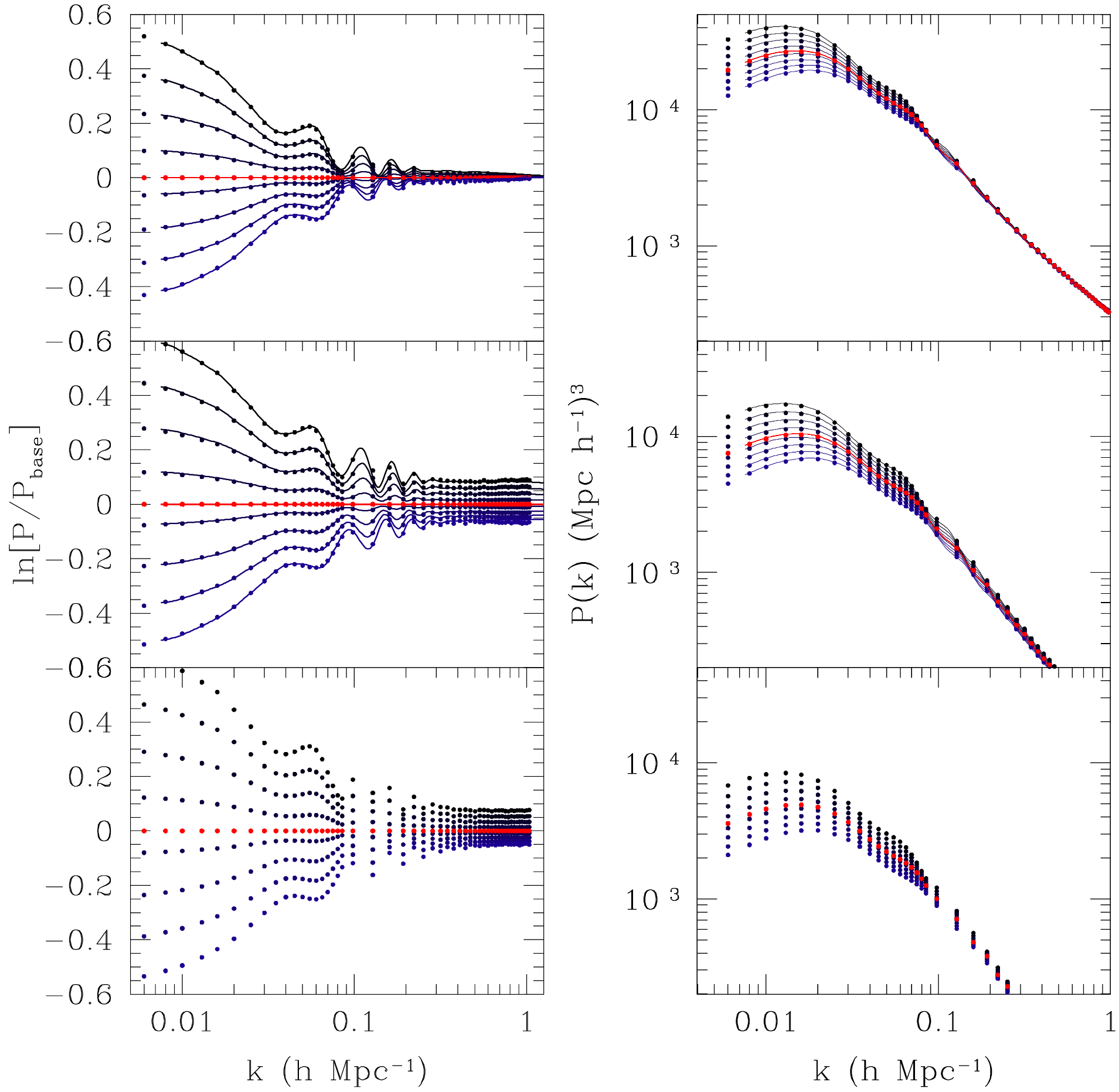}
        \caption{\small{Variations in the power spectrum at redshift $z=0$ (top row), $z=1$ (middle row), $z=2$ (bottom row). Parameter $\Omega_{\rm m}h^2$ is varied between its minimum and maximum value (see testing set range, Table~\ref{tab:train_set_priors}) while $\Omega_{\rm b} h^2, n_s, w, \sigma_8$ are fixed at their central values. $\sum m_\nu=0$ to facilitate comparison with the $h$-fixed version of the {\sc cosmic emulator} \citep{Lawrence10}. The left-hand panels show natural logarithm of the ratio of the power spectra with different $\Omega_{\rm m}h^2$ to the base power spectrum. The cosmological parameters for the base power spectrum are: ${\bf I}\equiv(0.135, 0.0225, 0.95, -1, 0.775, 0)$. The absolute power spectra are shown in the right-hand panels. Within each panel, the power spectra (from top to bottom) correspond to increasing values of $\Omega_{\rm m}h^2$. The predicted ratios using {\sc PkANN} (dotted) are within $0.2\%$ of the {\sc cosmic emulator}'s predictions (solid lines). At $z=2$, only {\sc PkANN} predictions are shown since the $h$-fixed {\sc cosmic emulator} is valid only for $z\leq1$.
    }}
    \label{fig:param1}
\end{figure*}

Having built the power spectrum interpolator, we now study the behavior of the power spectrum as a function of the cosmological parameters. Similar tests were performed by \cite{Heitmann13}. In Fig.~\ref{fig:param1}, we show variations in the power spectrum at redshift $z=0$ (top row), $z=1$ (middle row), $z=2$ (bottom row). At each redshift, $\Omega_{\rm m}h^2$ is varied between its minimum and maximum value (see parameter ranges for the testing set, in Table~\ref{tab:train_set_priors}) while $\Omega_{\rm b} h^2, n_s, w, \sigma_8$ are fixed at their central values.  We fix $\sum m_\nu=0$ since we want to compare our {\sc PkANN} predictions with the {\sc cosmic emulator}, which is not trained for cosmological models with massive neutrinos. The left-hand panels show natural logarithm of the ratio of the power spectra with different $\Omega_{\rm m}h^2$ to the base power spectrum. The base power spectrum corresponds to the central values: $\Omega_{\rm m} h^2=0.135, \Omega_{\rm b} h^2=0.0225, n_s=0.95, w=-1, \sigma_8=0.775$, with $\sum m_\nu=0$. The absolute power spectra are shown in the right-hand panels. Within each panel, the power spectra (from top to bottom) correspond to increasing $\Omega_{\rm m}h^2$. Higher $\Omega_{\rm m}h^2$ reduces the large-scale normalization of the power spectrum significantly. Accurate measurements of the power spectrum amplitude on large scales can help improve the constraints on $\Omega_{\rm m}h^2$. {\sc PkANN} predictions (dotted) agree well with the {\sc cosmic emulator} (solid lines).  Note that for redshift $z=2$, we only show {\sc PkANN} predictions since the $h$-fixed version of the {\sc cosmic emulator} \citep{Lawrence10} can make predictions only up to $z=1$.

\begin{figure*}
     \includegraphics[scale=0.8]{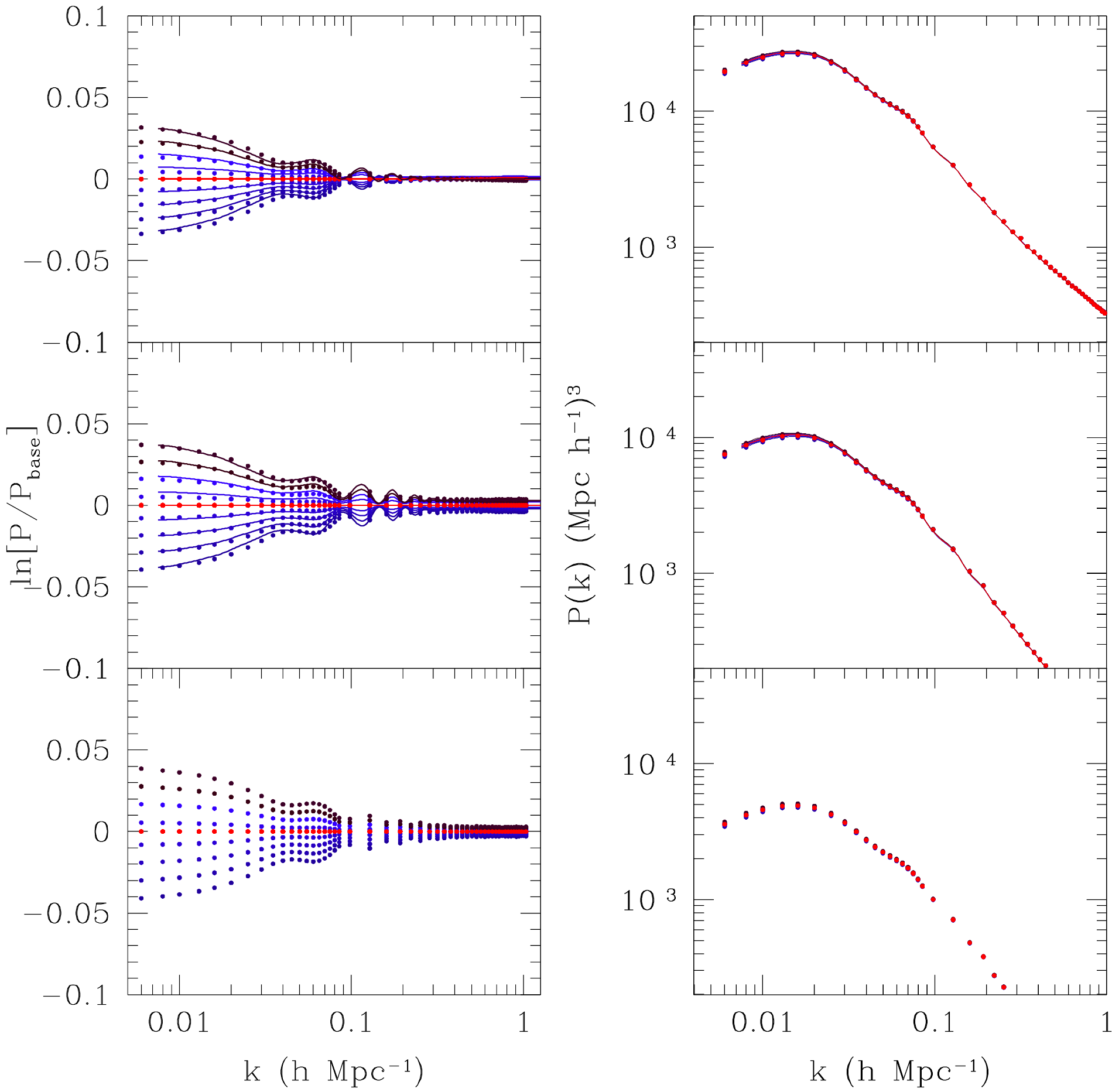}
        \caption{\small{Similar to Fig.~\ref{fig:param1}, but for a range of $\Omega_{\rm b}h^2$ values. Within each panel, the power spectra from bottom to top correspond to increasing $\Omega_{\rm b}h^2$ values.
    }}
    \label{fig:param31}
\end{figure*}

\begin{figure*}
     \includegraphics[scale=0.8]{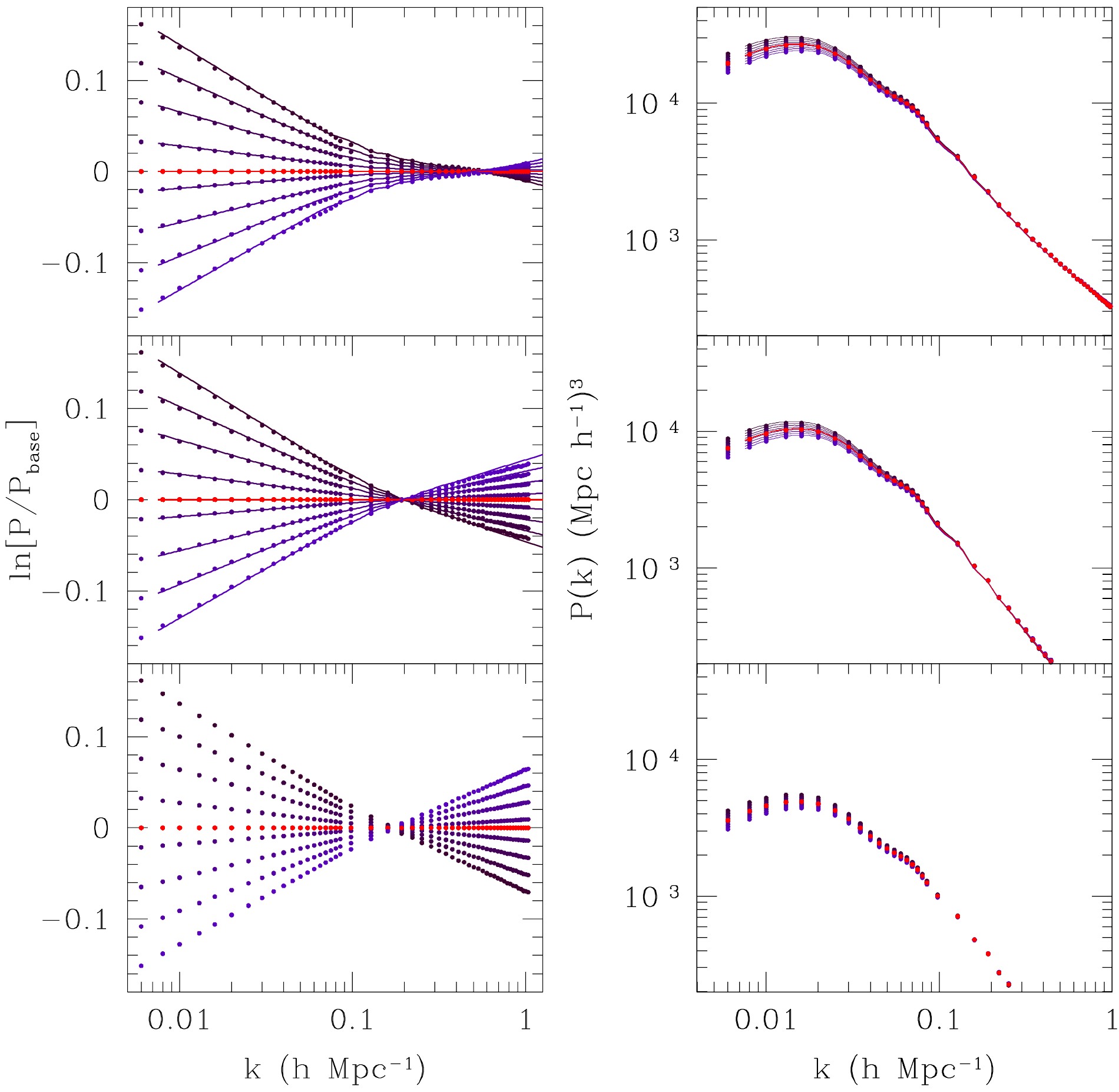}
        \caption{\small{Similar to Fig.~\ref{fig:param1}, but for a range of $n_s$ values. Within each panel, the power spectra from top to bottom correspond to increasing $n_s$ values.
    }}
    \label{fig:param61}
\end{figure*}

\begin{figure*}
     \includegraphics[scale=0.8]{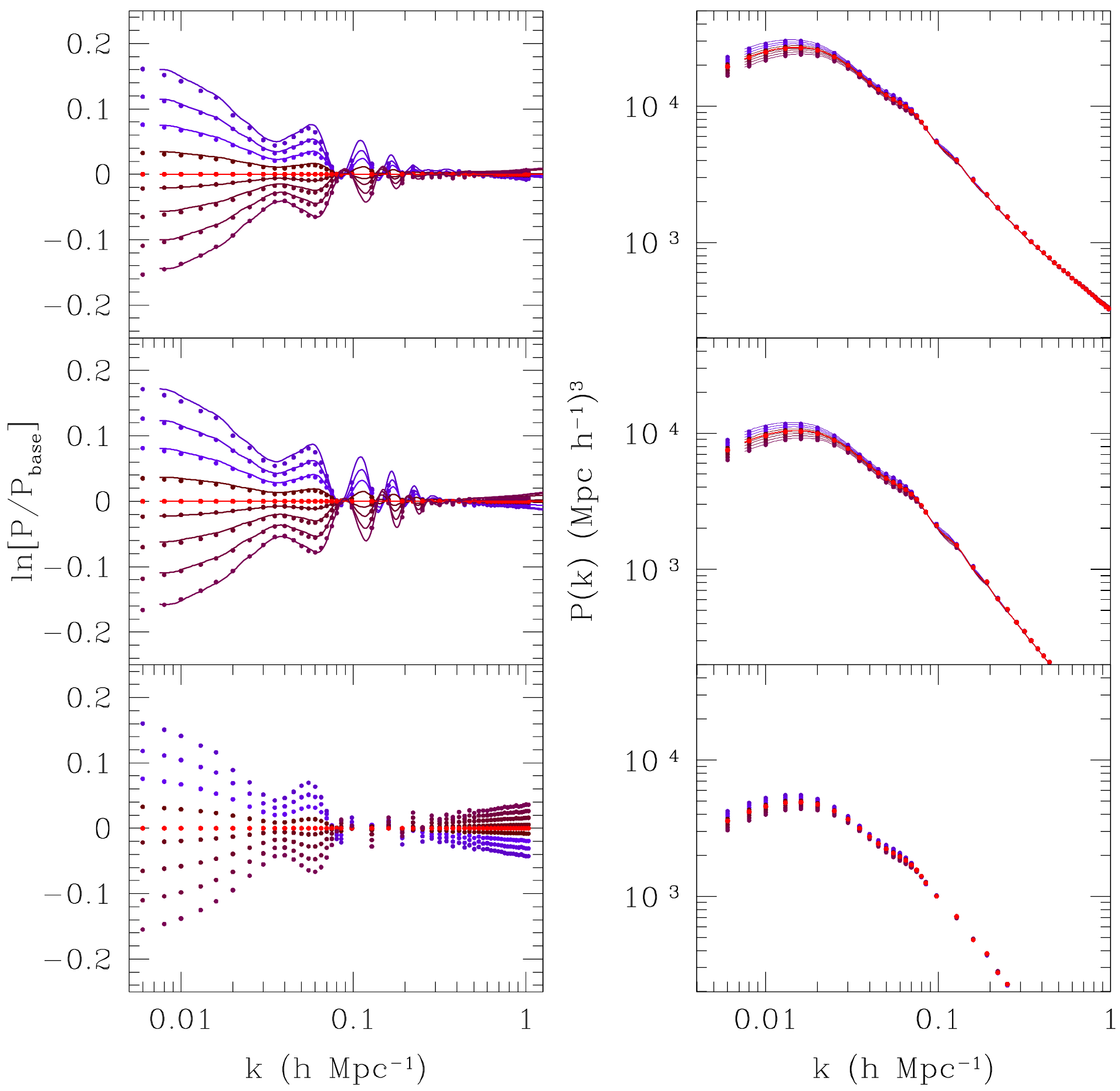}
        \caption{\small{Similar to Fig.~\ref{fig:param1}, but for a range of $w$ values. Within each panel, the power spectra from top to bottom correspond to increasing $w$ values.
    }}
    \label{fig:param91}
\end{figure*}

\begin{figure*}
     \includegraphics[scale=0.8]{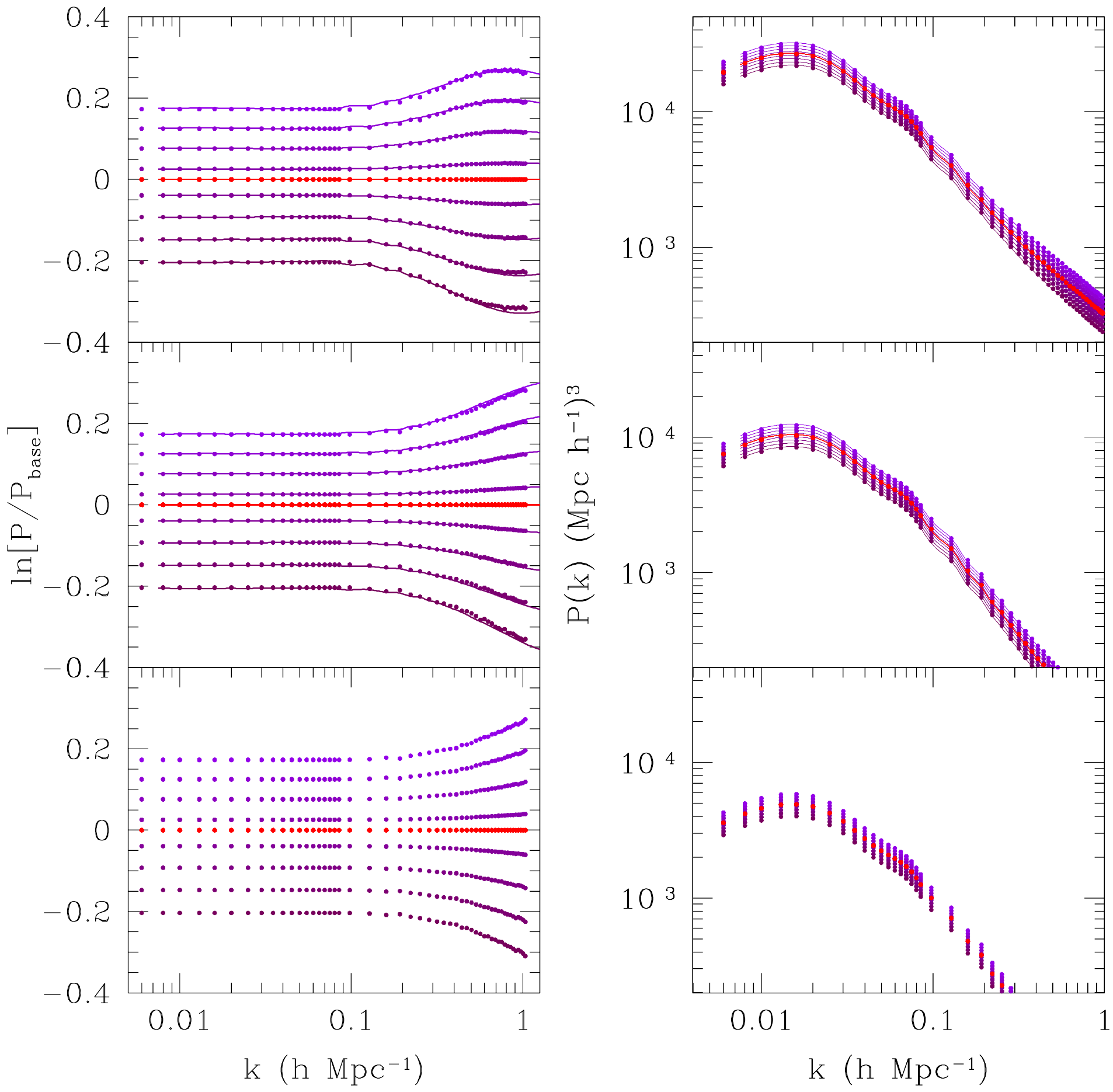}
        \caption{\small{Similar to Fig.~\ref{fig:param1}, but for a range of $\sigma_8$ values. Within each panel, the power spectra from bottom to top correspond to increasing $\sigma_8$ values.
    }}
    \label{fig:param121}
\end{figure*}

In Figs.~\ref{fig:param31} -- \ref{fig:param121}, we vary $\Omega_{\rm b} h^2, n_s, w$ and $\sigma_8$, respectively. The power spectra trends from minimum to maximum values are as follows: top to bottom ($n_s$ and $w$) and bottom to top ($\Omega_{\rm b} h^2$ and $\sigma_8$). At $z=0$, except $\sigma_8$, all other parameters affect the power spectrum predominantly on large scales ($\sim k<0.1\,h \textrm{Mpc}^{-1}$). Reducing uncertainties in the other parameters using small-scale data would be difficult unless one measures the power spectrum at higher redshifts where almost all parameters leave discernible imprints. Note that the power spectra converge around $k\sim0.1\,h \textrm{Mpc}^{-1}$ in the $\Omega_{\rm m} h^2, \Omega_{\rm b} h^2, n_s$ and $w$ plots. This is a direct consequence of our imposing the CMB constraint on the acoustic scale based on WMAP 7-year+BAO data. The acoustic scale is model dependent. Fixing its value to match observations requires adjusting the Hubble parameter $h$ accordingly. As we discussed in Section~\ref{sec:SIM}, given a cosmological model {\bf I}, we compute $h$ to satisfy the constraint ${\pi d_{ls}}/{r_s}=302.54$.

\begin{figure}
      \includegraphics[width= \columnwidth]{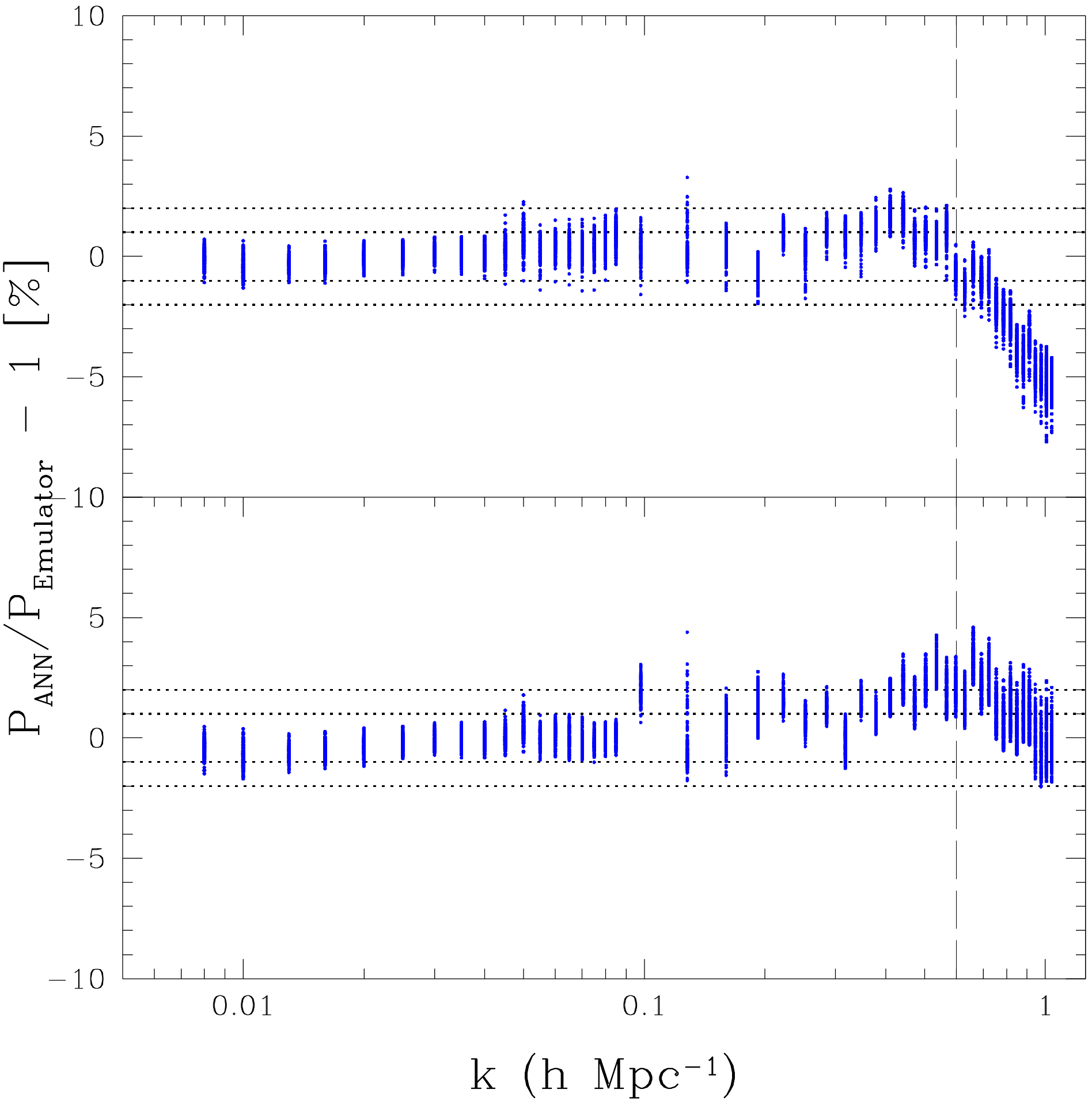}
        \caption{\small{The ratio of the matter power spectra at $z=0$ (upper panel) and $z=1$ (lower panel) computed using {\sc PkANN} and the $h$-fixed {\sc cosmic emulator} for all models of Section~\ref{sec:EXPLORE} as well as the 150 testing set cosmologies with $\sum m_\nu=0$. The two prediction schemes disagree at 5 per cent level out to $k\ltwid0.9\,h \textrm{Mpc}^{-1}$. Beyond $k=0.6\,h \textrm{Mpc}^{-1}$, {\sc PkANN} predictions fall off due to our use of unigrid simulations (see Section~\ref{sec:ERRORS} for discussion).
        }}
    \label{fig:ratio_pkann_coyote}
\end{figure}

In Fig.~\ref{fig:ratio_pkann_coyote}, we plot the ratio of the spectra at redshifts $z=0$ (upper panel) and $z=1$ (lower panel) computed using {\sc PkANN} and the $h$-fixed {\sc cosmic emulator}. The cosmologies considered are all models of Section~\ref{sec:EXPLORE} as well as the 150 testing set cosmologies with $\sum m_\nu=0$. At $z=0$, {\sc PkANN} matches the {\sc cosmic emulator}'s predictions to within $\pm2$ per cent up to $k\leq0.6\,h \textrm{Mpc}^{-1}$ and within $\pm5$ per cent up to $k\leq0.9\,h \textrm{Mpc}^{-1}$. The loss of power due to our use of $512^3$ unigrid simulations is clearly evident beyond $k=0.6\,h \textrm{Mpc}^{-1}$. At $z=1$, {\sc PkANN} is within $\pm5$ per cent over the scales considered.

\section{Conclusions} \label{sec:conclusions}

Machine-learning techniques offer unparalleled advantage in analyses of large datasets of the kind being generated by current and upcoming surveys. A brute force application of numerical simulations can consume millions of CPU-hours and may not be a feasible solution. Instead, by running a limited number of simulations, one can develop machine-learning schemes. These schemes can then be used to efficiently handle new and previously unseen data.

In this paper, we have introduced {\sc PkANN} -- the non-linear matter power spectrum interpolator. Using a manageable number of N-body simulations, we have successfully developed a neural network-based interpolating scheme that reconstructs the matter power spectrum over the parameter space spanning $3\sigma$ CL around the WMAP 7-year central values. Although {\sc PkANN} reproduces the input power spectrum at sub-percent level, its overall accuracy is limited by the accuracy of our N-body simulations. {\sc PkANN} (i) predicts matter power spectrum up to redshifts $z\le2$, (ii) is capable of handling spatially flat cosmological models with/without massive neutrinos, as well as dark energy models with $w\ne-1$ constant equation of state parameter, (iii) is accurate at 5 per cent level up to $k\leq0.9\,h \textrm{Mpc}^{-1}$ for models with $\sum m_\nu<0.5$ eV for all redshifts $z\leq2$, (iv) is accurate at 5 (10) per cent level for redshifts $z>1$ ($z\leq1$) for models with $\sum m_\nu>0.5$ eV.

The new generation of experiments, such as the DESI redshift maps, will measure matter density fluctuations with precision approaching $\sim1$ per cent level. Such unprecedented precision, while having the potential to refine constraints on various cosmological parameters, poses a tremendous challenge on theoretical predictions of the matter power spectrum. Baryon physics alters the power spectrum at $\sim20\%$ level at $k\approx10\,h \textrm{Mpc}^{-1}$. \cite{VanJoop11} have shown that AGN feedback reduces power relative to CDM-only simulations at per cent level at $k\approx0.4\,h \textrm{Mpc}^{-1}$. While the dark energy component in numerical simulations is usually assumed smooth and implemented only through its effects on the background evolution, \cite{AliFuzBouRasCouPSC2010} find that dark energy clustering leaves distinct imprints on the non-linear matter power spectrum. To extract any meaningful and unbiased information from current and upcoming data, such effects will need to be incorporated in N-body simulations and any fitting functions thereof. Although we did not consider a wide range of cosmological models such as the ones with non-zero spatial curvature, time-varying equation of state for dark energy or dark energy clustering, our ANN scheme can be readily extended for these cases by running extra simulations. The {\sc PkANN} package is freely available at \url{http://zuserver2.star.ucl.ac.uk/~fba/PkANN/PkANN.tar.gz}.

\noindent
\section{Acknowledgements}

This work is supported by the National Science Foundation through TeraGrid resources provided by the NCSA. The research leading to these results has received funding from the European Research Council under the European Community's Seventh Framework Programme (FP7/2007-2013 Grant Agreement no. 279954). FBA acknowledges the support of a Royal Society URF, and OL acknowledges a Royal Society Wolfson Research Merit Award and an Advanced ERC Grant. SA would like to thank Pier Stefano Corasaniti and Yann Rasera for useful discussion. We thank the referee for providing constructive comments and help in improving the contents of this paper.

\bibliographystyle{mn2e}
\bibliography{pkann2}

\renewcommand\thesection{\Alph{section}}
\setcounter{section}{0}
\renewcommand\thesubsection{\thesection.\arabic{subsection}}

\section{Appendix}
\label{sec:APPEND}

The following is based on the treatment presented in \cite{Bishop95}.

\subsection{PkANN Cost Function}
\label{sec:PkANN}

{\sc PkANN} is a single hidden-layer feed-forward network with sigmoid hidden nodes and linear output nodes. For training the {\sc PkANN} neural network to predict the matter power spectrum, we consider a training set consisting of cosmological models for which we have full information about the non-linear matter power spectra $P_{\mathrm{nl}}$ (computed from N-body simulations) as a function of scale $k$ and redshift $z$, as well as the underlying cosmological parameters: ${\bf I}\equiv(\Omega_{\rm m} h^2, \Omega_{\rm b} h^2, n_s, w, \sigma_8, \sum m_\nu)$. The joint likelihood of getting the set of matter power spectra $\{P_{\mathrm{nl}}(z;{\bf I}_t)\}$ for all cosmologies ${\bf I}_t$ in the training set is
\bea
 \label{eq:LIKE1} \nonumber
 \L \left[\{P_{\mathrm{nl}}(z;{\bf I}_t)\}\right] &=& \prod_{t=1}^T {\it p} [P_{\mathrm{nl}}(z;{\bf I}_t)]\\
                                                              &=& \prod_{t=1}^T {\it p} [P_{\mathrm{nl}}(z|{\bf I}_t)] \,\, {\it p} [{\bf I}_t],
\eea
where ${\it p}[P_{\mathrm{nl}}(z|{\bf I}_t)]$ is to be interpreted as the conditional probability of getting spectrum $P_{\mathrm{nl}}(z)$ {\it given} cosmology ${\bf I}_t$, while ${\it p} [{\bf I}_t]$ is the unconditional probability that the cosmological parameters ${\bf I}$ take a particular setting of ${\bf I}_t$. The index $t$ runs over all cosmologies ${\bf I}_t$ in the training set. We can take the product of the individual probabilities since each model ${\bf I}_t$ is drawn independently from the cosmological parameter space.

The weights {\bf w} of the {\sc PkANN} network are chosen (iteratively during network training) so as to minimize the negative logarithm of the likelihood $\L$ (which is equivalent to maximizing $\L$),
\bea
\label{eq:LIKE2}
\chi^2  \,\,=\,\, -\ln \L &=& \sum_{t=1}^T \ln {\it p} [P_{\mathrm{nl}}(z|{\bf I}_t)] + \sum_{t=1}^T \ln {\it p} [{\bf I}_t].
\eea

If the power spectrum is sampled at different values of scale $k$ (the $k$-modes being represented by the set $\{k\}\, h \textrm{Mpc}^{-1}$), we can write ${\it p}[P_{\mathrm{nl}}(z|{\bf I}_t)]$ as
\bea
\label{eq:LIKE3}
{\it p}[P_{\mathrm{nl}}(z|{\bf I}_t)] &=& \prod_{{k_i\in\{k\}}} {\it p} [P_{\mathrm{nl}}(k,z|{\bf I}_t)],
\eea
where the product $k_i$ is over all the scales that form the set $\{k\}\, h \textrm{Mpc}^{-1}$, and we have assumed that $P_{\mathrm{nl}}(k,z|{\bf I}_t)$ have independent distributions.

To suppress sampling uncertainties in the power spectrum $P_{\mathrm{nl}}(k,z|{\bf I}_t)$, the numerical simulation code is run multiple times with different seeds while keeping the underlying cosmological model ${\bf I}_t$ fixed. Assuming $P_{\mathrm{nl}}(k,z|{\bf I}_t)$ has Gaussian distribution about the true power spectrum $P^{\mathrm{Tr}}_{\mathrm{nl}}(k,z|{\bf I}_t)$ with variance $\sigma^2$, we can write the probability that a numerical run would give $P_{\mathrm{nl}}(k,z|{\bf I}_t)$ as
\bea
\label{eq:LIKE4}
{\it p} [P_{\mathrm{nl}}(k,z|{\bf I}_t)] &=& \frac{1}{{(2\pi\sigma^2)}^{1/2}} e^{\frac{-\left[P^{\mathrm{Tr}}_{\mathrm{nl}}(k,z|{\bf I}_t)-P_{\mathrm{nl}}(k,z|{\bf I}_t)\right]^2}{2\sigma^2}}.
\eea

N-body codes give larger variance $\sigma^2$ on scales comparable to the simulation volume since the density field on these scales can only be sampled fewer times. However, to simplify the {\sc PkANN} training algorithm, in Eq.~\ref{eq:LIKE4} we have assumed that the variance $\sigma^2$ is independent of the scale $k$ and model ${\bf I}_t$. 
 
Since the aim of developing {\sc PkANN} is to model the true spectrum $P^{\mathrm{Tr}}_{\mathrm{nl}}(k,z|{\bf I}_t)$ by making an optimal choice for the network weights ${\bf w}$, we replace $P^{\mathrm{Tr}}_{\mathrm{nl}}(k,z|{\bf I}_t)$ in Eq.~\ref{eq:LIKE4} by the ANN prediction $P_{\mathrm{nl}}^{\rm ANN}(k,z|{\bf w,I}_t)$ to get
\bea
\label{eq:LIKE5}
{\it p} [P_{\mathrm{nl}}(k,z|{\bf I}_t)] = \frac{1}{{(2\pi\sigma^2)}^{1/2}} e^{\frac{-\left[P_{\mathrm{nl}}^{\rm ANN}(k,z|{\bf w,I}_t)-P_{\mathrm{nl}}(k,z|{\bf I}_t)\right]^2}{2\sigma^2}}.
\eea
Inserting Eq.~\ref{eq:LIKE5} into Eq.~\ref{eq:LIKE3}, we get

\bea
\label{eq:LIKE6} \nonumber
{\it p}[P_{\mathrm{nl}}(z|{\bf I}_t)] &=& \frac{1}{{(2\pi\sigma^2)}^{N_{out}/2}} \\
&& e^{\frac{-\sum_{{k_i\in\{k\}}} \left[P_{\mathrm{nl}}^{\rm ANN}(k,z|{\bf w,I}_t)-P_{\mathrm{nl}}(k,z|{\bf I}_t)\right]^2}
{2\sigma^2}},
\eea
where $N_{out}$ is the number of $k$-modes in the set $\{k\}$. Remember that, by construction, $N_{out}$ is also the number of nodes in the output layer of the {\sc PkANN} network. Using Eq.~\ref{eq:LIKE6}, we can now write Eq.~\ref{eq:LIKE2} as
\bea
\label{eq:LIKE7} \nonumber
\chi^2({\bf w}) \!\!\! &=& \!\!\! \frac{1}{2\sigma^2} \sum_{t=1}^T \sum_{{k_i\in\{k\}}} \left[P_{\mathrm{nl}}^{\rm ANN}(k,z|{\bf w,I}_t)-P_{\mathrm{nl}}(k,z|{\bf I}_t)\right]^2 \\
                         && -\; T \ln \left[ \frac{1}{{(2\pi\sigma^2)}^{N_{out}/2}} \right] +\; \sum_{t=1}^T \ln {\it p} [{\bf I}_t].
\eea

We can drop the terms that do not depend on the weights ${\bf w}$, since these terms merely scale $\chi^2({\bf w})$ without altering its location in the weight-space. Thus, the cost function for the purposes error minimization can be written as
\bea
\label{eq:LIKE8}
\chi^2({\bf w}) = \frac{1}{2} \sum_{t=1}^T \sum_{{k_i\in\{k\}}} \left[P_{\mathrm{nl}}^{\rm ANN}(k,z|{\bf w,I}_t)-P_{\mathrm{nl}}(k,z|{\bf I}_t)\right]^2.
\eea
Remember that the matter power spectrum is a function of scale $k\, (h \textrm{Mpc}^{-1})$. We sample the matter spectrum at discreet values between $0.006\,h \textrm{Mpc}^{-1} \leq k \leq 1\,h \textrm{Mpc}^{-1}$ and assign the sampled spectrum to the output nodes of the neural network. The discreet values of scale $k$ form the set $\{k\}\, h \textrm{Mpc}^{-1}$. In Eq.~\ref{eq:LIKE8}, the sum $k_i$ is over all the nodes in the output layer, with each node sampling the matter power spectrum at some specific scale, $k\, (h \textrm{Mpc}^{-1})$. $P_{\mathrm{nl}}(k,z|{\bf I})$ is the non-linear matter power spectrum for the specific cosmology ${\bf I}$, computed using N-body simulations. Given the weights ${\bf w}$, $P_{\mathrm{nl}}^{\rm ANN}(k,z|{\bf w,I})$ is the ANN's predicted power spectrum for the ${\bf I}$th cosmology. In our fitting procedure, we work with the ratio of the non-linear to linear power spectrum, namely $R(k,z)\equiv P_{\mathrm{nl}}(k,z)/P_{\mathrm{lin}}(k,z)$, where $P_{\mathrm{lin}}(k,z)$ is calculated using {\sc camb} \citep{Lewis00}. As such, weighing Eq.~\ref{eq:LIKE8} by $P_{\mathrm{lin}}(k,z)$ gives,
\bea
\label{eq:chisq} \nonumber
\chi^2({\bf w}) \!\!\!\!\!\! &=& \!\!\!\!\!\! \frac{1}{2} \sum_{t=1}^T\sum_{k_i\in\{k\}} \left[\frac{P_{\mathrm{nl}}^{\rm ANN}(k,z|{\bf w,I}_t)-P_{\mathrm{nl}}(k,z|{\bf I}_t)} {P_{\mathrm{lin}}(k,z|{\bf I}_t)}\right]^2 \\
 \!\!\!\!\!\! &=& \!\!\!\!\!\! \frac{1}{2} \sum_{t=1}^T\sum_{k_i\in\{k\}} \left[R^{\rm ANN}(k,z|{\bf w,I}_t)-R(k,z|{\bf I}_t)\right]^2.
\eea
The ratio $R(k,z)$ is a flatter function and gives better performance, particularly at higher redshifts where the ratio tends to 1. Given the weights ${\bf w}$, $R^{\rm ANN}(k,z|{\bf w,I})$ in Eq.~\ref{eq:chisq} is the network's prediction of the ratio $R(k,z|{\bf I})$ for the specific cosmology ${\bf I}$. The predicted non-linear spectrum $P_{\mathrm{nl}}^{\rm ANN}(k,z|{\bf w,I})$ is recovered by multiplying $R^{\rm ANN}(k,z|{\bf w,I})$ by the corresponding linear spectrum $P_{\mathrm{lin}}(k,z|{\bf I})$.

In Eq.~\ref{eq:chisq}, optimizing the weights ${\bf w}$ so as to minimize $\chi^2$({\bf w}) generates an ANN that predicts the power spectrum very well for the specific cosmologies in the training set. However, such a network might not make accurate predictions for cosmologies $\it{not}$ included in the training set. This usually indicates (i) an overly simple network architecture (very few hidden layer nodes), (ii) very sparsely or poorly sampled parameter space and/or (iii) a highly complex non-linear mapping that actually over-fits to the noise on the training dataset. In order to generate smoother network mappings that generalize better when presented with new cosmologies that are not part of the training set, a penalty term $\chi^2_{Q}({\bf w})$ is added to the cost function $\chi^2({\bf w})$,
\be
\label{eq:penalty}
\chi^2_{Q}({\bf w}) = \frac{\xi}{2} || {\bf w} ||^2,
\ee
where $|| {\bf w} ||^2$ is the quadratic sum of all the weights. The penalty term $\chi^2_{Q}({\bf w})$ prevents the network weights from becoming too large during the training process by penalizing in proportion to their sum. The regularization parameter $\xi$ controls the degree of regularization (smoothing) of a network's predictions. After having initialized $\xi$, its value is re-estimated during the training process iteratively. For the update formula, see Appendix \ref{sec:PENALTY}. For its derivation, see \cite{Bishop95}.

Thus, the overall cost function which is presented to the ANN for minimization with respect to the weights $\bf w$ is,
\bea
\label{eq:cost} \nonumber
\chi^2_{C}({\bf w}) \!\!\! &=& \!\!\! \frac{1}{2}\sum_{t=1}^T\sum_{k_i\in\{k\}} \left[R^{\rm ANN}(k,z|{\bf w,I}_t)-R(k,z|{\bf I}_t)\right]^2\\
&& +\; \frac{\xi}{2} || {\bf w} ||^2.
\eea

\subsection{Quasi-Newton Method}
\label{sec:QNM}

Quasi-Newton method, used for finding stationary points (local maxima and minima) of a function, assumes that the function can be approximated by a quadratic in the region around a stationary point. Taylor expanding the {\sc PkANN} cost function $\chi^2_C({\bf w})$ (see Eq.~\ref{eq:cost}) around some point ${\bf w}_0$ in the weight space and retaining terms up to second-order, we get
\bea
\label{eq:Q1} \nonumber
\chi^2_C({\bf w}) &=& \chi^2_C({\bf w}_0) + ({\bf w - w}_0)^T {\bf g}_{{\bf w}_0} \\
&& \!\!\!\!\! +\; \frac{1}{2} ({\bf w - w}_0)^T {\bf H}_{{\bf w}_0} ({\bf w - w}_0),
\eea
where the superscript $T$ stands for the transpose and ${\bf g}_{{\bf w}_0}$ is defined to be the gradient of $\chi^2_C$ evaluated at ${\bf w}_0$,
\bea
\label{eq:Q2}
{\bf g}_{{\bf w}_0} \equiv \bigtriangledown \chi^2_C \big|_{{\bf w}_0}.
\eea
${\bf H}_{{\bf w}_0}$ is a symmetric $N_W \times N_W$ Hessian matrix (evaluated at ${\bf w}_0$) with elements
\bea
\label{eq:Q3}
H_{ij}\big|_{{\bf w}_0} \equiv \frac{\partial^2 \chi^2_C}{\partial w_i \partial w_j}\bigg|_{{\bf w}_0},
\eea
where $N_W$ (see Eq.~\ref{eq:Nodes}) is the total number of nodes in the network. Note that in Eq.~\ref{eq:Q3}, instead of referencing the weights by the relevant nodes they connect to, for the sake of clarity we refer to the weights with a single subscript running from 1 to $N_W$.

Taking the gradient of Eq.~\ref{eq:Q1} gives the local approximation for the gradient itself,
\bea
\label{eq:Q4}
{\bf g_{w}} &=& {\bf g}_{{\bf w}_0} + {\bf H}_{{\bf w}_0} ({\bf w - w}_0).
\eea

To find the stationary point around ${\bf w}_0$, one sets ${\bf g_{w}}$ in Eq.~\ref{eq:Q4} to zero, thereby giving the {\it Newton step},
\bea
\label{eq:Q5}
{\bf w} &=& {\bf w}_0 - {\bf H}^{-1}_{{\bf w}_0} {\bf g}_{{\bf w}_0}.
\eea

Since the cost function $\chi^2_C({\bf w})$ is not an exact quadratic function, the Newton step of Eq.~\ref{eq:Q5} does not point to the local minimum around ${\bf w}_0$. As such, we apply Eq.~\ref{eq:Q5} iteratively, and if the Hessian matrix is positive definite (\ie all of its eigenvalues are positive), then each successive Newton step moves closer to the local minimum. If the initial choice of the weights ${\bf w}$ happens to be around a local maximum of $\chi^2_C({\bf w})$, then the Hessian matrix is not positive definite and the cost function may increase with each Newton step.

One can apply some modifications to the Newton method that guarantee convergence towards a local minimum, irrespective of the initial choice of the weights. Instead of taking a step in the {\it Newton direction} (${-\bf H}^{-1} {\bf g}$), one proceeds in a {\it quasi-Newton direction} (${-\bf G {\bf g}}$),
\bea
\label{eq:Q6}
{\bf w} &=& {\bf w}_0 - \lambda_{{\bf w}_0} {\bf G}_{{\bf w}_0} {\bf g}_{{\bf w}_0},
\eea
where matrix ${\bf G}$ represents an approximation to the inverse of the Hessian ${\bf H}^{-1}$, and $\lambda$ is the size of the step taken along the quasi-Newton direction ${-\bf G {\bf g}}$. The step size $\lambda$ is allowed to vary with each iteration to the weights. Its value is determined by proceeding in the direction ${-\bf G {\bf g}}$ until the minimum of the cost function is found along ${-\bf G {\bf g}}$. Thus, in Eq.~\ref{eq:Q6}, $\lambda_{\bf {w_0}}$ is such that the gradient of $\chi^2_C$ at ${\bf w}$ (namely, ${\bf g_w}$) vanishes along the direction ${-\bf G}_{{\bf w}_0} {\bf g}_{{\bf w}_0}$,
\bea
\label{eq:Q7}
\left( {-\bf G}_{{\bf w}_0} {\bf g}_{{\bf w}_0} \right)^T {\bf g_{w}} &=& 0.
\eea

The quasi-Newton algorithm involves taking a a series of steps $\tau$ of Eq.~\ref{eq:Q6}, which can be written as
\bea
\label{eq:Q8}
{\bf w}_{\tau+1} &=& {\bf w}_\tau - \lambda_{\bf {w_\tau}} {\bf G}_{{\bf w}_\tau} {\bf g}_{{\bf w}_\tau},
\eea
with the step size $\lambda_{{\bf w}_\tau}$ for the $\tau$th step being such that
\bea
\label{eq:Q9}
\left( {-\bf G}_{{\bf w}_\tau} {\bf g}_{{\bf w}_\tau} \right)^T {\bf g}_{{\bf w}_{\tau+1}} &=& 0.
\eea

At each step of the algorithm, ${\bf G}$ is constructed to be positive definite, ensuring that the direction ${-\bf G {\bf g}}$ proceeds towards a local minimum of the cost function. To construct ${\bf G}$, we use the Broyden--Fletcher--Goldfarb--Shanno (BFGS) method (see Appendix \ref{sec:CFH}).

\subsection{PkANN Cost Function Gradient}
\label{sec:CFG}

The overall cost function which is presented to the ANN for minimization with respect to the weights $\bf w$ is given by Eq.~\ref{eq:cost}.

We now derive the expression for its derivative with respect to the weights ${\bf w}$. {\sc PkANN}'s network architecture is $N_{in}$ :  $N_1$ :  $N_{out}$ with two layers of adaptive weights. The first layer of weights $w_{ji}$ connect the input layer nodes $(x_0,x_1,...,x_i)$ to the hidden nodes $(z_1,...,z_j)$. Note that the hidden bias node activation $z_0$ is permanently fixed at 1 and therefore, does not receive any connections from the input layer. The activation of each hidden node is $z_j\equiv g(a_j)$, taking as its argument
\bea
\label{eq:aj2}
a_j &=& \sum_{i=0}^{N_{in}} w_{ji}x_i,
\eea
where the sum is over all input nodes $i$ (including the input bias) sending connections to the $j$th hidden node (barring the hidden bias node).

{\sc PkANN}'s hidden nodes have sigmoidal activations $g(a_j) = 1/[1 +\rm exp(-a_j)]$. The second layer of weights $w_{kj}$ connect the hidden nodes $(z_0,z_1,...,z_j)$ to the network outputs $(y_1,...,y_k)$. The output nodes have linear activations $y_k=a_k$, with $a_k$ being the weighted sum of all hidden nodes,
\bea
\label{eq:ak2}
a_k &=& \sum_{j=0}^{N_1} w_{kj}z_j.
\eea

PkANN has two layers of adaptive weights and we will consider the cost function derivatives separately for the two layers.

\subsubsection{Gradient w.r.t. $\!\!\!$ First Layer Weights}

Taking the gradient of Eq.~\ref{eq:cost} with respect to a first layer weight $w_{ji}$, we get 
\bea
\label{eq:G1} \nonumber
\frac{\partial \left[\chi^2_C({\bf w})\right]}{\partial w_{ji}} \!\!\!\!\!\! &=& \!\!\!\!\!\! \sum_{t,{\{k\}}} \! \left[R^{\rm ANN}(k,z|{\bf w,I}_t)-R(k,z|{\bf I}_t)\right] \! \frac{\partial R^{\rm ANN}}{\partial w_{ji}} \\
&&  \!\!\!\!\!\! +\; \xi w_{ji}.
\eea
Since $R^{\rm ANN}(k,z|{\bf w,I}_t)=a(k,z|{\bf w,I}_t)$ (see Eq.~\ref{eq:ak2}) for the output nodes, we get
\bea
\label{eq:G2} \nonumber
\frac{\partial \left[\chi^2_C({\bf w})\right]}{\partial w_{ji}} \!\!\!\!\!\! &=& \!\!\!\!\!\! \sum_{t,{\{k\}}} \! \left[R^{\rm ANN}(k,z|{\bf w,I}_t)-R(k,z|{\bf I}_t)\right] \! \frac{\partial a_k^t}{\partial w_{ji}} \\
&&  \!\!\!\!\!\! +\; \xi w_{ji},
\eea
where we have introduced the shorthand notation $a_k^t \equiv a(k,z|{\bf w,I}_t)$. Using Eq.~\ref{eq:ak2} for $a_k$ together with sigmoidal activation for  $z_j$, we get
\bea
\label{eq:G3} \nonumber
\frac{\partial a_k^t}{\partial w_{ji}} & = & \sum_{j^\prime=0}^{N_1} w_{kj^\prime} \frac{\partial z_{j^\prime}^t}{\partial w_{ji}} \\
& = & \sum_{j^\prime=0}^{N_1} w_{kj^\prime} \frac{\partial g(a_{j^\prime}^t)}{\partial a_{j^\prime}^t} \frac{\partial a_{j^\prime}^t}{\partial w_{ji}}.
\eea

For sigmoidal activation functions, it is straightforward to show that 
\bea
\label{eq:G4}
\frac{\partial g(a_j^t)}{\partial a_j^t} &=& g(a_j^t)\left(1-g(a_j^t)\right).
\eea

Inserting Eq.~\ref{eq:G4} into Eq.~\ref{eq:G3}, we get
\bea
\label{eq:G5}
\frac{\partial a_k^t}{\partial w_{ji}} & = & \sum_{j^\prime=0}^{N_1} w_{kj^\prime} g_{j^\prime}^t \left(1-g_{j^\prime}^t\right) \frac{\partial a_{j^\prime}^t}{\partial w_{ji}}.
\eea

Differentiating Eq.~\ref{eq:aj2} with respect to a first layer weight $w_{ji}$, we get
\bea
\label{eq:G6} \nonumber
\frac{\partial a_{j^\prime}^t}{\partial w_{ji}} &=& \sum_{i^\prime=0}^{N_{in}} x_{i^\prime}^t \frac{\partial w_{{j^\prime}{i^\prime}}}{\partial w_{ji}} \\
&=& \sum_{i^\prime=0}^{N_{in}} x_{i^\prime}^t \delta_{ii^{\prime}} \delta_{jj^{\prime}}  \;=\; x_i^t \delta_{jj^{\prime}}.
\eea

Inserting Eq.~\ref{eq:G6} into Eq.~\ref{eq:G5}, we get
\bea
\label{eq:G7} \nonumber
\frac{\partial a_k^t}{\partial w_{ji}} & = & \sum_{j^\prime=0}^{N_1} w_{kj^\prime} g_{j^\prime}^t \left(1-g_{j^\prime}^t\right) x_i^t \delta_{jj^{\prime}} \\
& = & w_{kj} g_{j}^t \left(1-g_{j}^t\right) x_i^t.
\eea

From Eqs.~\ref{eq:G2} and \ref{eq:G7}, we get our final equation for the derivative of the {\sc PkANN}\ cost function with respect to the first layer of adaptive weights $w_{ji}$ to be
\bea
\label{eq:G8} \nonumber
\frac{\partial \left[\chi^2_C({\bf w})\right]}{\partial w_{ji}} \!\!\!\!\!\! &=& \!\!\!\!\!\! \sum_{t,{\{k\}}} R^{\rm ANN}(k,z|{\bf w,I}_t) \; w_{kj} g_{j}^t \left(1-g_{j}^t\right) x_i^t \\
&& \!\!\!\!\!\!\!\!\! - \sum_{t,{\{k\}}} R(k,z|{\bf I}_t) \; w_{kj} g_{j}^t \left(1-g_{j}^t\right) x_i^t + \xi w_{ji}.
\eea

\subsubsection{Gradient w.r.t. $\!\!\!$ Second Layer Weights}

Taking the gradient of Eq.~\ref{eq:cost} with respect to a second layer weight $w_{kj}$, we get 
\bea
\label{eq:G9} \nonumber
\frac{\partial \left[\chi^2_C({\bf w})\right]}{\partial w_{kj}} \!\!\!\!\!\! &=& \!\!\!\!\!\! \sum_{t,{\{k^\prime\}}} \!\! \left[R^{\rm ANN}(k^\prime,z|{\bf w,I}_t)-R(k^\prime,z|{\bf I}_t)\right] \! \frac{\partial R^{\rm ANN}}{\partial w_{kj}} \\
&& \!\!\!\!\!\! +\; \xi w_{kj}.
\eea
Since $R^{\rm ANN}(k^\prime,z|{\bf w,I}_t)=a(k^\prime,z|{\bf w,I}_t)$ (see Eq.~\ref{eq:ak2}) for the output nodes, we get
\bea
\label{eq:G10} \nonumber
\frac{\partial \left[\chi^2_C({\bf w})\right]}{\partial w_{kj}} \!\!\!\!\!\! &=& \!\!\!\!\!\! \sum_{t,{\{k^\prime\}}} \!\! \left[R^{\rm ANN}(k^\prime,z|{\bf w,I}_t)-R(k^\prime,z|{\bf I}_t)\right] \! \frac{\partial a_{k^\prime}^t} {\partial w_{kj}} \\
&& \!\!\!\!\!\! +\; \xi w_{kj},
\eea
where as before, we use the shorthand notation $a_{k^\prime}^t \equiv a(k^\prime,z|{\bf w,I}_t)$. From Eq.~\ref{eq:ak2}, we get
\bea
\label{eq:G11} \nonumber
\frac{\partial a_{k^\prime}^t} {\partial w_{kj}} & = & \sum_{j^\prime=0}^{N_1}  \frac{\partial w_{k^\prime j^\prime}} {\partial w_{kj}} z_{j^\prime}^t \\
& = & \sum_{j^\prime=0}^{N_1}  \delta_{kk^\prime} \delta_{jj^\prime} z_{j^\prime}^t \;=\; \delta_{kk^\prime} z_j^t
\eea

Inserting Eq.~\ref{eq:G11} into Eq.~\ref{eq:G10}, we get our final equation for the derivative of the {\sc PkANN}\ cost function with respect to the second layer of adaptive weights $w_{kj}$ to be
\bea
\label{eq:G12} \nonumber
\frac{\partial \left[\chi^2_C({\bf w})\right]}{\partial w_{kj}} \!\!\!\!\!\! &=& \!\!\!\!\!\!\!\! \sum_{t,{\{k^\prime\}}} \!\! \left[R^{\rm ANN}(k^\prime,z|{\bf w,I}_t)-R(k^\prime,z|{\bf I}_t)\right] \! \delta_{kk^\prime} z_j^t \\ \nonumber
&& \!\!\!\!\!\! +\; \xi w_{kj} \\ \nonumber
\!\!\!\!\!\! &=& \!\!\!\!\!\! \sum_t \! \left[R^{\rm ANN}(k,z|{\bf w,I}_t)-R(k,z|{\bf I}_t)\right] \! z_j^t + \xi w_{kj} \\
\eea

For any choice of weights ${\bf w}$, the network output vector $R^{\rm ANN}(k,z|{\bf w,I}_t)$ is determined for each cosmology ${\bf I}_t$ in the training set, by progressing sequentially through the network layers, from inputs to outputs, calculating the activation of each node. Having calculated the activations and network outputs for all cosmologies, it is straightforward to compute the derivatives in Eqs.~\ref{eq:G8} and \ref{eq:G12}.

\subsection{BFGS Approximation for Inverse-Hessian Matrix}
\label{sec:CFH}

In order to minimize the {\sc PkANN} cost function $\chi^2_C({\bf w})$ (see Eq.~\ref{eq:cost}) with respect to the weights ${\bf w}$, the weights are first randomly initialized to ${\bf w}_0$ and then updated iteratively using Eq.~\ref{eq:Q8}.

Updating the weights involves estimating ${\bf G}$ -- an approximation to the inverse Hessian matrix ${\bf H}^{-1}$. The inverse Hessian ${\bf H}^{-1}$ evaluated at ${\bf w_0}$ is approximated by a $N_W \times N_W$ identity matrix (\ie ${\bf G}_{{\bf w}_0}={\bf I}$). Following our discussion in Appendix \ref{sec:QNM}, the weight vector is updated to ${\bf w}_1$ as
\bea
\label{eq:H1}
{\bf w}_1 &=& {\bf w}_0 - \lambda_{{\bf w}_0} {\bf g}_{{\bf w}_0}
\eea
by stepping a distance $\lambda_{\bf {w_0}}$ in the quasi-Newton direction ${-\bf g}_{{\bf w}_0}$. Note that the gradient ${\bf g}_{{\bf w}_0}$ is computed using Eqs.~\ref{eq:G8} and \ref{eq:G12}. The step size $\lambda_{{\bf w}_0}$ is such that the gradient of $\chi^2_C$ at ${\bf w}_1$ (namely, ${\bf g}_{{\bf w}_1}$) vanishes along the direction ${-\bf g}_{{\bf w}_0}$,
\bea
\label{eq:H2}
-{\bf g}^T _{{\bf w}_0} {\bf g}_{{\bf w}_1} &=& 0.
\eea

To make any further updates in the weight space, one needs to evaluate ${\bf H}^{-1}_{{\bf w}_1}$. The inverse Hessian, being a $N_W \times N_W$ matrix, can be computationally expensive to calculate exactly for networks with $N_W \gtwid 1000$ connections. We employ the BFGS method to approximate ${\bf H}^{-1}_{{\bf w}_1}$ by ${\bf G}_{{\bf w}_1}$. In general, for the $(\tau+1)$ step, the approximation ${\bf G_{w_{\tau+1}}}$ is
\bea
\label{eq:H3} \nonumber
{\bf G}_{{\bf w}_{\tau+1}} = {\bf G}_{{\bf w}_\tau} \\
&& \!\!\!\!\!\!\!\!\!\!\!\!\!\!\!\!\!\!\!\!\!\!\!\!\!\!\!\!\!\!\!\! +\; \frac{1}{S_1} \left[ \left( 1 + \frac{S_2}{S_1} \right) {\bf a a}^T- {\bf a b}^T {\bf G}_{{\bf w}_\tau} - {\bf G}_{{\bf w}_\tau} {\bf b a}^T \right],
\eea
where we use the following definitions for the vectors (${\bf a}$ and ${\bf b}$) and the scalars ($S_1$ and $S_2$),
\bea
\label{eq:H4} \nonumber
{\bf a} &=& {\bf w}_{\tau+1} - {\bf w}_\tau \\ \nonumber
{\bf b} &=& {\bf g}_{{\bf w}_{\tau+1}} - {\bf g}_{{\bf w}_\tau} \\ \nonumber
S_1 &=& {\bf a}^T {\bf b} \\
S_2 &=& {\bf b}^T {\bf Gb}
\eea

At each step, the BFGS method makes increasingly more accurate approximations for ${\bf G}$. Moreover, since ${\bf G}$ is positive definite (by construction), the $\chi^2_C({\bf w})$ minimization algorithm is guaranteed to converge to a local minimum.

\subsection{Regularization Parameter $\xi$}
\label{sec:PENALTY}

In situations where the training data is noisy, controlling the complexity of a network is crucial to avoid overfitting and underfitting issues. An overly complex network may fit the noise in the training data. On the other hand, a very simple network may not be able to capture the signal in a dataset, leading to underfitting. Both overfitting and underfitting lead to models with low predictive performance. One of the methods employed to regularize the complexity of a neural network is to train the network by minimizing a cost function that includes a penalty term $\chi^2_{Q}({\bf w})$ (\eg see Eq.~\ref{eq:penalty}).

Small (large) values of the regularization parameter $\xi$ lead to complex (simple) networks. Since the optimum value for $\xi$ is not known a priori, its value is initialized randomly, and updated iteratively by the cost minimization algorithm.

Here, we only present the updating rule for $\xi$. For its derivation, refer \cite{Bishop95}. The {\sc PkANN} cost function (Eq.~\ref{eq:cost}) can be written as

\be
\label{eq:P1}
\chi^2_{C}({\bf w}) \!\!=\!\! \beta \! \left( \! \frac{1}{2} \! \sum_{t,{\{k\}}} \! \left[R^{\rm ANN}(k,z|{\bf w,I}_t)\!-\!R(k,z|{\bf I}_t)\right]^2 \!\!+\! \frac{\alpha}{2\beta} || {\bf w} ||^2 \! \right)
\ee

where $\alpha$ and $\beta$ are the regularization parameters with $\xi \equiv \alpha / \beta$ and $\beta=1$. For the purposes of cost minimization, the overall scale factor $\beta$ is irrelevant and the degree of regularization depends only on the ratio $\xi \equiv \alpha / \beta$. For networks where the number of training patterns $N_T$ far exceeds the number of weights $N_W$, \cite{Bishop95} derives the following updating rules for $\alpha$ and $\beta$,
\bea
\label{eq:P3}
\alpha_{\tau+1} &=& \frac{N_W}{|| {\bf w}_\tau ||^2} \\
\beta_{\tau+1} &=& \frac{N_T}{\chi^2({\bf w}_\tau)},
\eea
where $\chi^2({\bf w})$ (see Eq.~\ref{eq:chisq}) is the sum of squares of residuals for the training data. Thus, we update $\xi$ as
\bea
\label{eq:P4}
\xi_{\tau+1} &=& \frac{N_W}{N_T} \frac{\chi^2({\bf w}_\tau)}{|| {\bf w}_\tau ||^2}.
\eea
From Eq.~\ref{eq:P4}, we see that for sufficiently complex networks ($N_W>>1$) with lots of training data ($N_T>>N_W$), the parameter $\xi<<1$. It shows that underfitting and overfitting issues can be avoided by simply choosing network architectures that satisfy conditions: (i) $N_W>>1$ and (ii) $N_T>>N_W$. However, both these conditions can put tremendous load on the computing resources. In situations where the computing time is at a premium, a penalty term is used to achieve a balance between computing load and desired prediction accuracy of the neural network.

\end{document}